\documentclass[aps,pre,showpacs,twocolumn,amsmath,amssymb,amsfonts,eqsecnum]{revtex4-1}
\usepackage{graphicx}
\usepackage{amssymb}
\usepackage{epstopdf}
\usepackage{hyperref}
\usepackage{float}
\usepackage{color}
\usepackage{ulem}

\begin{document}

\title{Protein viscoelastic dynamics: a model system  }

\author{Craig Fogle}
\email{cfogle@physics.ucla.edu}
\author{Joseph Rudnick}
\email[]{jrudnick@physics.ucla.edu}
\affiliation{Department of Physics and Astronomy, UCLA, Box 951547, Los Angeles, CA, 90095-1547}
\author{David Jasnow}
\email{jasnow@pitt.edu}
\affiliation{Department of Physics and Astronomy, University of Pittsburgh, Pittsburgh, PA, 15260}

\date{\today}

\begin{abstract}
A model system  inspired by recent experiments on the dynamics of a folded protein under the influence of a sinusoidal force \cite{zocchi1,zocchi2,zocchi4,zocchi3} is investigated and found to replicate  many of the  response characteristics of such a system. The essence of the model is  a strongly over-damped  oscillator described by a harmonic restoring force for small displacements that reversibly yields to stress under sufficiently large displacement.  This simple dynamical system also reveals unexpectedly rich behavior---exhibiting a series of dynamical transitions and analogies with equilibrium thermodynamic phase transitions. The effects of noise and of inertia are briefly considered and  described.
\end{abstract}


\maketitle

\section{Introduction}

Investigations by Zocchi and collaborators \cite{zocchi1,zocchi2,zocchi4,zocchi3}  of the dynamics of folded proteins under the influence of a sinusoidally modulated force provide insight into the mechanical response of those molecules to external forces. These investigations probe both the interactions and the dynamics associated with conformational changes in proteins. The latter aspect of the work promises to enhance our understanding of the action of proteins, since conformational adjustments, particularly substantial alterations in the tertiary structure of these molecules, are central to their action in key biological settings \cite{whitford,physcell}.

In the experimental system studied by Zocchi \textit{ et. al.} a collection of guanylate kinase enzymes tether 20 nm gold nano-particles to a planar gold substrate through Cysteins that are mutagenically introduced into those enzymes. An oscillating electrophoretic force, generated by applying an AC voltage between the gold substrate and a parallel electrode, drives the charged gold nano-particles; the motion of those nano-particles is then measured with the use of evanescent wave scattering. The large number of gold nano particles in the sample that participate in the motion allow for the detection of displacements that are considerably smaller than thermal motion would predict. Those displacements map directly onto the deformation of the enzymes.

\begin{figure}[htbp]
\begin{center}
\includegraphics[width=2.5in]{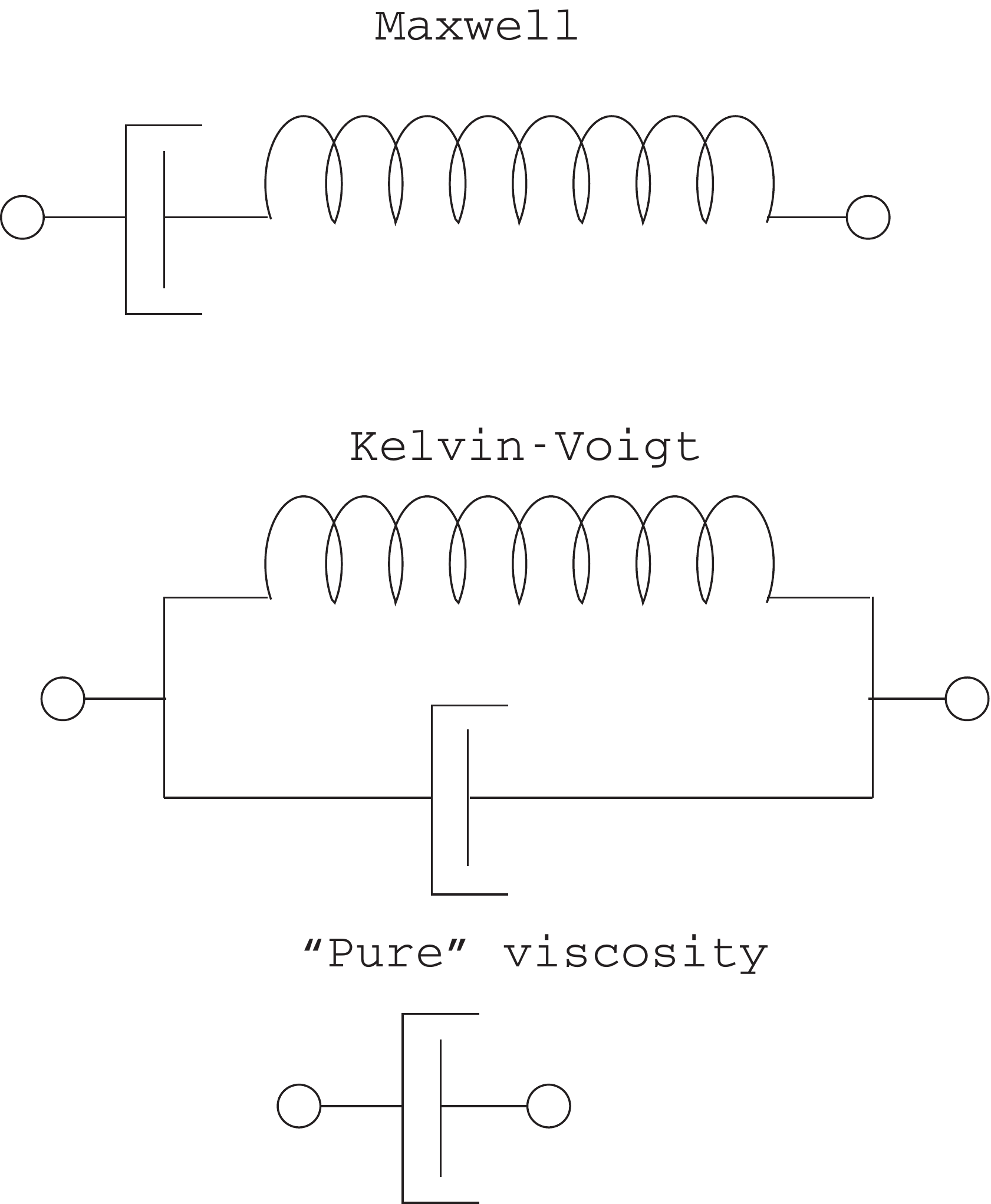}
\caption{Two simple models for a viscoelastic system, along with a model for a system governed entirely by viscosity. The springs have spring constants equal to $k$, and the viscous coefficient for the dash pots is $\gamma$. }
\label{fig:models}
\end{center}
\end{figure}

A key finding that emerges from these investigations is the existence of a relatively abrupt crossover as the driving force increases, from elastic response to a response that is dominantly viscous. This crossover is described as a ``viscoelastic transition.'' Reference \cite{zocchi3} discusses the nature of the transition and describes two response regimes. First, when the displacement is small, the most accurate underlying model of the enzyme appears to be a Hookean spring, for which the equation of motion is simply $x(t) = f(t)/k$, where $f(t)$ is the applied sinusoidal force.
This representation of the protein characterizes its deformation in terms of a single collective coordinate, $x(t)$, and neglects both inertial and dissipative effects. On the other hand, for sufficiently large driving force, the motion is described in terms of a Maxwellian model of a dissipative system \cite{Lakes}, schematically displayed in Fig. \ref{fig:models}.
The equations of motion that govern this system are
\begin{eqnarray}
\gamma \frac{dy(t)}{dt} & = & f(t) \label{eq:meq1} \\
x(t) &=& y(t) + \frac{\gamma}{k} \frac{dy(t)}{dt} \label{eq:meq2}
\end{eqnarray}
where $x(t)$ once again tracks the deformation of the enzyme, which is directly related to the displacement of the gold nano particle, while $y(t)$ is an ``internal" degree of freedom of the system. (A further time differentiation of Eq.(\ref{eq:meq2}) renders it in the familiar form relating strain rate to the stress and its rate of change.) According to \cite{zocchi3}, the mechanism that drives the transition is an underlying energy versus displacement curve that changes from quadratic---i.e.,  harmonic---at low displacement, to linear at higher displacements. The precise nature of this form of potential energy curve is not expanded on in that reference.

Inspired by the above results, we have explored the dynamical response of a somewhat different---but unified---depiction of the driven protein system. Our approach is based on a Kelvin-Voigt model for a viscoelastic substance, as shown in Fig. \ref{fig:models}. The spring in this system is not, however, strictly Hookean. Rather the energy of the spring as a function of displacement, $V(x)$, which gives rise to a restoring force $F(x)$, is schematically displayed in Fig. \ref{fig:enforce}. In our initial analysis, we ignore inertial effects, under the assumption that the system is heavily over-damped. We find that the system displays an unexpectedly rich range of behavior, including symmetry breaking---and restoration---dynamical phase transitions, as well as  noise driven rounding and  ``switching" in bi-stability and many features reminiscent of equilibrium thermodynamic phase transitions, such as
spinodals and multicriticality.  Preliminary investigations of the effects of inertia on the dynamical equations reveal an even richer range.

There is at least one precedent for a dynamical transition in a driven system of the type that is explored here. The Suzuki-Kubo equation describes the behavior of a mean field version of the Ising model with dissipative dynamics in which the spin variables are driven by a magnetic field with sinusoidal time dependence \cite{sk}. At low temperatures, the response to the driving field undergoes a dynamical transition, from an oscillation about an equilibrium ferromagnetic state at small amplitude of the drive to an oscillation at larger drive amplitudes centered about spin magnitude equal to zero \cite{TandO}. This dynamical transition can be either continuous or first order, depending on the temperature. However, the physics underlying that model differs fundamentally from the phenomena explored here.

The remainder of this paper is organized as follows. In Sec. \ref{sec:model} the basic model is introduced and the deterministic, noise free, dynamical phase diagram is displayed summarizing the basic behavior. In Sec. \ref{sec:analytical}, to provide additional insight, we provide some analytical results for the model with a piecewise continuous restoring force. In the following section, Sec. \ref{sec:compliances}, we make contact with typical dynamical responses, as measured in experiments on viscoelastic materials in general and in the experiments of Zocchi et al. in particular. For the basic over-damped deterministic case we discuss the nature of the transitions and compare the analysis to that of standard mean field theory for thermodynamic phase transitions in Sec. \ref{sec:transnature}. Section \ref{sec:inertia} contains a preliminary investigation into the effects of inertia, while the effect of noise on the transitions and response functions is discussed in Sec. \ref{sec:noise} via Langevin over-damped dynamics and a master equation. Concluding remarks follow, including a short discussion of the consequences of a restoring force without the symmetry shown in Fig. \ref{fig:enforce}. Appendices contain some details on our ``standard model,'' a comparison to mean field thermodynamics and a brief commentary on an alternate version of the restoring force.

\section{Model, dynamical phase diagram, and characterization of response} \label{sec:model}

\begin{figure}[htbp]
\begin{center}
\includegraphics[width=3in]{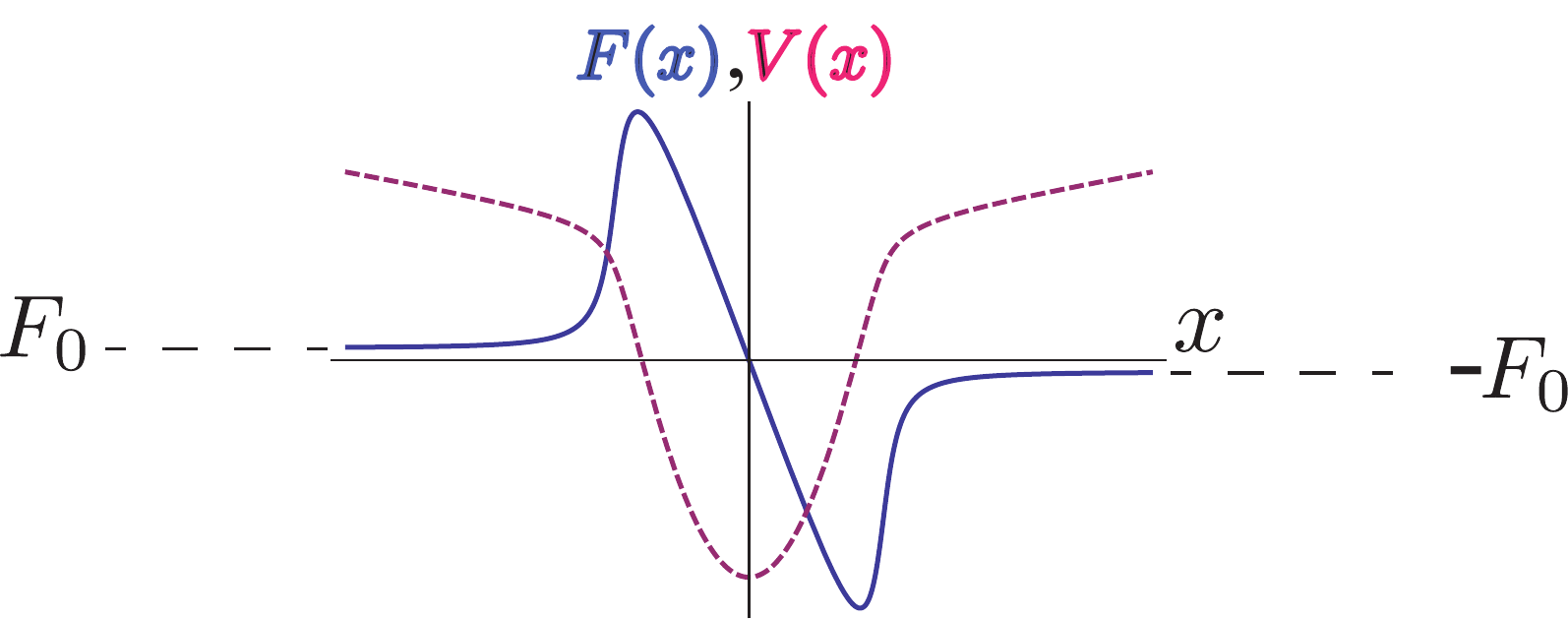}
\caption{The potential energy, $V(x)$, (red, dashed) and associated restoring force, $F(x)$, (blue, solid) that we assume in our model. The force approaches $\pm F_0$ for large $x$, as indicated in the figure.}
\label{fig:enforce}
\end{center}
\end{figure}

Our model is a highly simplified depiction of a folded protein, in which we single out a collective degree of freedom that we assume dominates the response of the molecule to an external force. As a caricature of its much more complex structure, we assume a system described by the Kelvin-Voigt model of viscoelasticty \cite{Lakes,Findley}, as shown in Fig. \ref{fig:models}, which, as we will see, appears natural for a driven, over-damped, nonlinear oscillator. The spring in this model can be thought of as a gross simplification of the Tirion model for protein interactions \cite{tirion}, with the caveat that our coordinate, $x$, is a collective one, while the Tirion model replaces the various interactions between the actual constituents of a protein by harmonic
bonds between idealized point-like entities. In addition, we assume the possibility of the ``cracking'' of this protein under sufficient external stress \cite{Ansari,proteinquakes,[{See also Anderson \textit{et. al.}: Chapter 8 of }] Noy} through the reversible detachment of the spring. This means that the harmonic potential energy expression
holds only in a limited range of values of $x$. According to the model we adopt, at sufficiently large values of $|x|$ the potential energy and associated restoring force depart from strict Hookean form
to approach either a constant energy (and, correspondingly, no restoring force) or a fixed, constant restoring force corresponding to a linear, rather than quadratic, energy. Figure \ref{fig:enforce} is a schematic depiction of a particular version of such a force and associated energy function. Inside a region centered at the origin, $x=0$, the potential is nearly harmonic and the force is approximately Hookean. Outside this central region the force approaches constant values $\pm F_0$, and the associated confining potential is linear. The precise form of the restoring force utilized in our calculations is described in Appendix \ref{app:restore}.

We assume initially that the motion of our system is highly over-damped, so that the time rate of change of the displacement coordinate $x$ is directly proportional to the force generating change in the system. Given the restoring force shown in Fig. \ref{fig:enforce} and a periodic external force, the equation of motion takes the form
\begin{equation}
\frac{dx(t)}{dt} = \frac{1}{\gamma} \left[ F(x(t)) + A \sin( \omega t)\right] \label{eq:eom1}
\end{equation}
where the parameter $\gamma$ encodes viscous effects.

\begin{figure}[htbp]
\begin{center}
\includegraphics[width=3in]{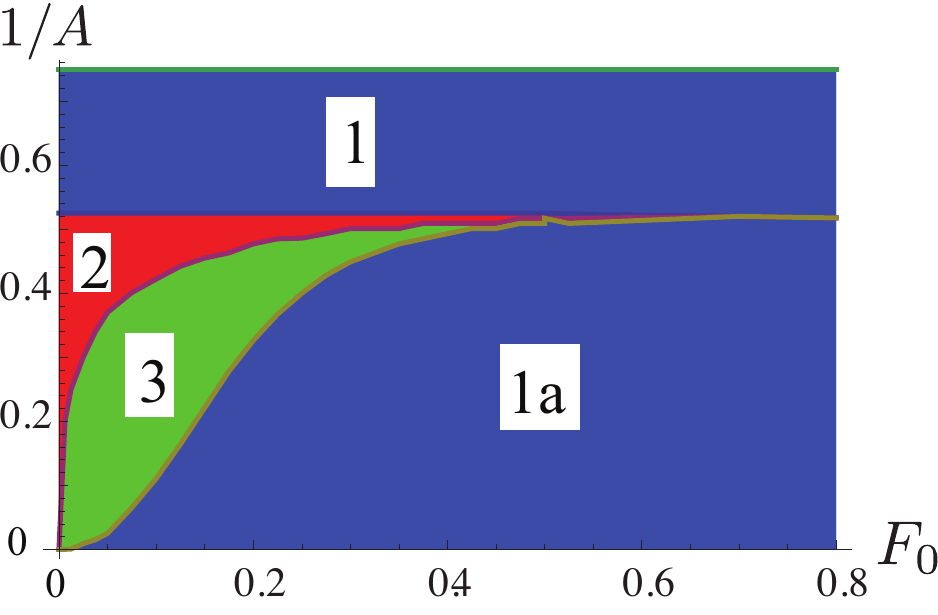}
\caption{Dynamical phase diagram of the types of  response to a sinusoidal drive exhibited by
the equation of motion (\ref{eq:eom1}) with restoring force $F(x)$ as shown in Fig. \ref{fig:enforce}. The horizontal axis is the absolute value of the asymptotes of the restoring force at large $|x|$. The vertical axis is the inverse of the drive amplitude, $A$. The parameter $\alpha$ controlling the transition from Hooke's law to constant restoring force (see Eqs. (\ref{eq:restore1})--(\ref{eq:restore6}) and Fig. \ref{fig:3Forces}) has been set equal to 0.01. The frequency of the drive, $\omega$, is equal to 1/2, and the viscosity parameter, $\gamma$, is set equal to one. }
\label{fig:phasediagrampic2}
\end{center}
\end{figure}

Figure \ref{fig:phasediagrampic2} indicates the resulting behavior when the drive amplitude, $A$, and $F_0$, the asymptotic absolute strength of the confining force, are scanned.
In Regions 1 and 1a, color-coded blue, the steady state periodic solution, is symmetric about the origin. In Region 2, color-coded red, the dynamically stable solution is skewed, either to the right or the left of the origin. As it turns out, skewed solutions appear in pairs; see Appendix \ref{app:pairs}. There is also a symmetric steady state solution that is, however, dynamically unstable. In Region 3, color-coded green, three dynamically stable, steady state solutions exist, one symmetric and the other two skewed. In addition, there are two skewed, \textit{dynamically unstable}, steady state solutions. Finally, a region complementing Region 3, separating Region 1 from Region 2, is not shown as it is exceedingly narrow and beyond the resolution in Fig. \ref{fig:phasediagrampic2}.

\begin{figure}[htbp]
\begin{center}
\includegraphics[width=3in]{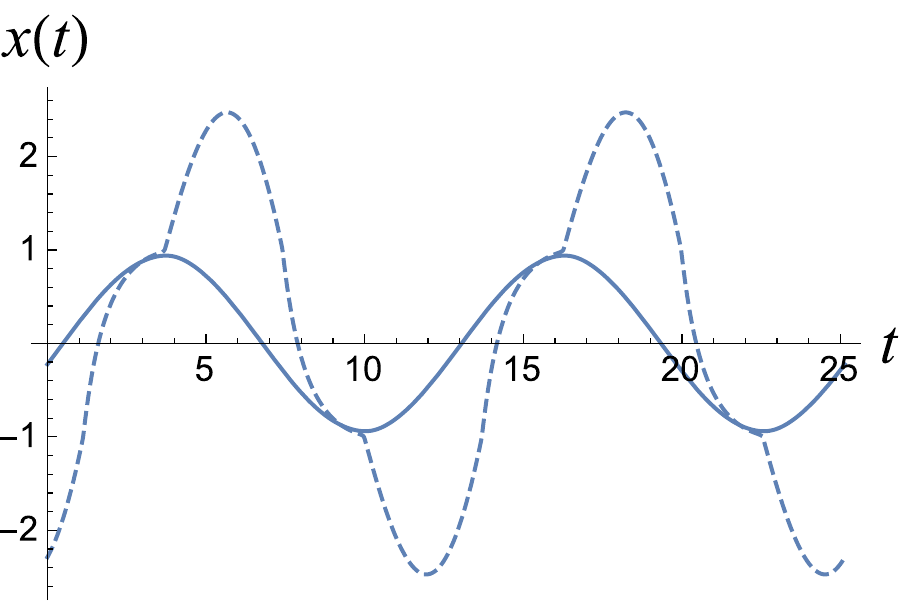}
\caption{The steady state response in Region 1 (solid curve) and in Region 1a (dashed curve). Both are dynamically stable. The inverse of the drive amplitude, $1/A$, is 0.53 for the Region 1 response and 0.51 for the Region 1a response. The asymptotes of the restoring force outside the harmonic regime are $\pm F_0$ with $F_0=0.6$, and the value of the spring constant in the Hooke's law region is $k=2$. The frequency of the driving force is $\omega =0.5$. The transition between a harmonic restoring force and a force equal to $\pm F_0$ occurs at $|x| \simeq 1$. The restoring force is given in Appendix \ref{app:restore} with parameter $\alpha =0.01$. }
\label{fig:1and1a}
\end{center}
\end{figure}

\begin{figure}[htbp]
\begin{center}
\includegraphics[width=3in]{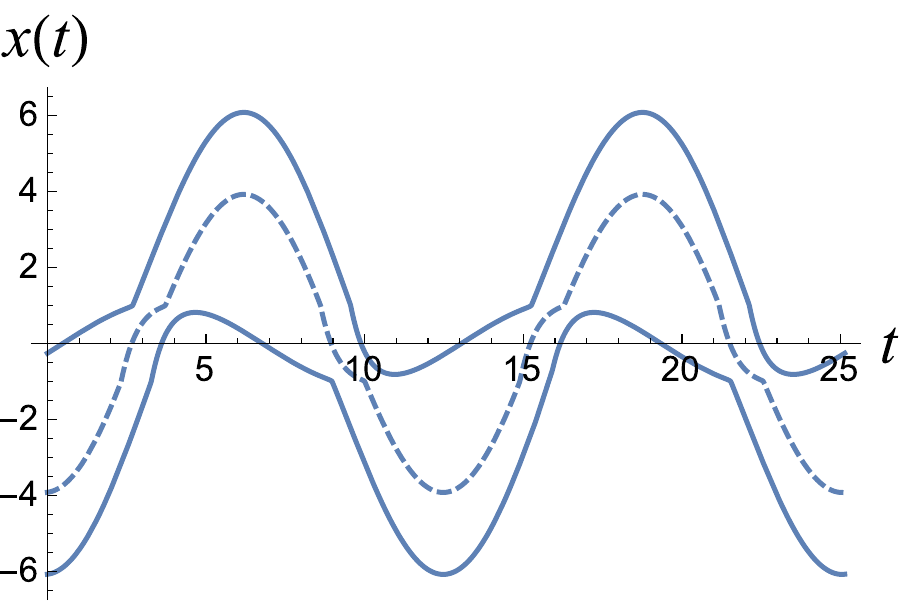}
\caption{The three steady state responses in Region 2. The solid, skewed, curves are dynamically stable and the dashed, symmetric, curve is dynamically unstable. The quantity $F_0$ is 0.1 and $1/A=0.45$. All other parameters are the same as in Fig. \ref{fig:1and1a}. }
\label{fig:2}
\end{center}
\end{figure}

\begin{figure}[htbp]
\begin{center}
\includegraphics[width=3in]{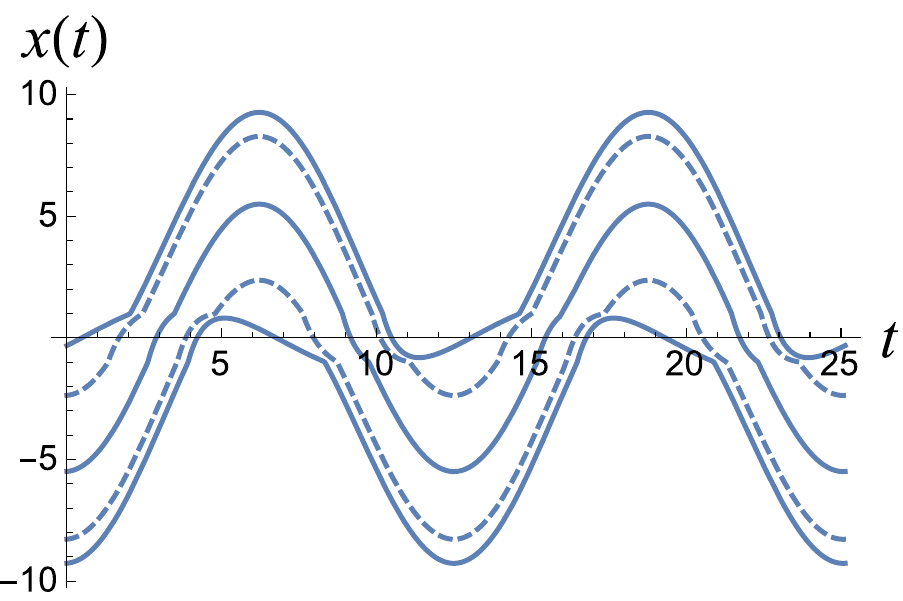}
\caption{The five steady state responses in Region 3. The solid symmetric curve and the two solid skewed curves are dynamically stable, while the two skewed curves shown dashed are dynamically unstable. The quantity $F_0$ is 0.1, and the inverse of the drive amplitude, $1/A$, is 0.35. All other parameters are the same as in Figs. \ref{fig:1and1a} and \ref{fig:2}.}
\label{fig:3}
\end{center}
\end{figure}

Figures \ref{fig:1and1a}--\ref{fig:3} illustrate the (steady state) solutions characteristic of the four regions in Fig. \ref{fig:phasediagrampic2}. The time interval shown corresponds to two periods of the driving force. The parameters $A$ and $F_0$ corresponding to the region of the phase diagram are provided in the captions, while $k=2$, corresponding to the spring constant in the harmonic regime, and $\omega=0.5$  for the driving force are kept constant in these figures. The friction constant $\gamma = 1$ sets the time scale. 
Given the very different properties of the steady states in the four regions,  the boundaries between the regions are necessarily sharp. We will return to the nature of the transitions that take place as those boundaries are traversed.

For the time being, it is worthwhile to consider the two solutions for $x(t)$ shown in Fig. \ref{fig:1and1a}. They differ in a few respects. First, although the driving force strengths differ by less than 5\%, the displacement amplitudes are quite different. Second, the curve describing the solution in Region 1 is nearly sinusoidal (in fact, nearly in phase with the drive, which goes as $\sin (\omega t)$, while the steady state solution for Region 1a is somewhat distorted. Furthermore, the latter solution is well out of phase with respect to the drive. As we will see, the steady state solution in Region 1 is close to elastic---i.e., nondissipative---while the the behavior in Region 1a is strongly dissipative.

\section{Analytical solution in a limiting case}  \label{sec:analytical} 
To gain a perspective on the nature of solutions of Eq. (\ref{eq:eom1}), we consider
the limiting assumption of a sharp break between the regime in which the restoring force is strictly linear and the range of the displacement variable $x$ in which it is considerably gentler. If the restoring force outside the harmonic regime \textit{ vanishes}, the function $F(x)$ has the form
\begin{equation}
F(x) = \left\{ \begin{array}{ll} -kx & |x| <x_0 \\ 0 & |x| \ge x_0 \end{array} \right. \label{eq:forcelim1}
\end{equation}
The solutions to
Eq. (\ref{eq:eom1}) in the two regimes are
\begin{eqnarray}
x_i(t) &=& Be^{-kt/\gamma} \nonumber \\ && - \frac{A}{\sqrt{\gamma^2 \omega^2 +k^2}} \cos \left( \omega t + \arctan \frac{k}{ \gamma \omega} \right) \label{eq:inside} \\ x_e(t) & = & B^{\prime} -\frac{A}{\gamma \omega} \cos ( \omega t) \, ,\label{eq:outside}
\end{eqnarray}
where the subscripts $i$ and $e$ refer to behavior in the ``interior'' region $|x|<x_0$ and the ``exterior'' region $|x| \ge x_0$. A complete solution to the equation of motion requires adjusting the coefficients $B$ and $B^{\prime}$ so as to match $x_i(t)$ and $x_e(t)$ at the boundary between the two regimes.

When the frequency of the driving force is small, the characters of the ``exterior'' and ``interior'' responses differ fundamentally. In the exterior region, the steady state velocity, $v_e(t) = dx_e(t)/dt$,  is in phase with the driving force, which means that the drive's energy feeds optimally into viscous damping. On the other hand, when $\gamma \omega \ll k$, $dx_i(t)/dt$ is almost ninety degrees out of phase with the driving force, so the response is nearly the same as at static equilibrium, in that the two terms in square brackets in (\ref{eq:eom1}) nearly cancel, and the dissipation is quite small. In the regime of low frequency response, we can roughly characterize the dynamics in the exterior region as \textit{viscous} and the behavior in the interior region as \textit{elastic}. It is therefore reasonable to expect that as the driving amplitude, $A$, increases, the response will evolve from elastic to viscous.

However, as strongly indicated by the phase diagram shown above, the change from elastic to viscous response entails an abrupt transition in the dynamical behavior. This transition occurs near the drive amplitude for which the coordinate $x$ visits the region outside the harmonic regime (in fact, it occurs at a slightly lower drive amplitude).  At this point skewed solutions to the equation of motion appear and, in fact, coexist with solutions remaining entirely within the harmonic regime.

\begin{figure}[htbp]
\begin{center}
\includegraphics[width=3in]{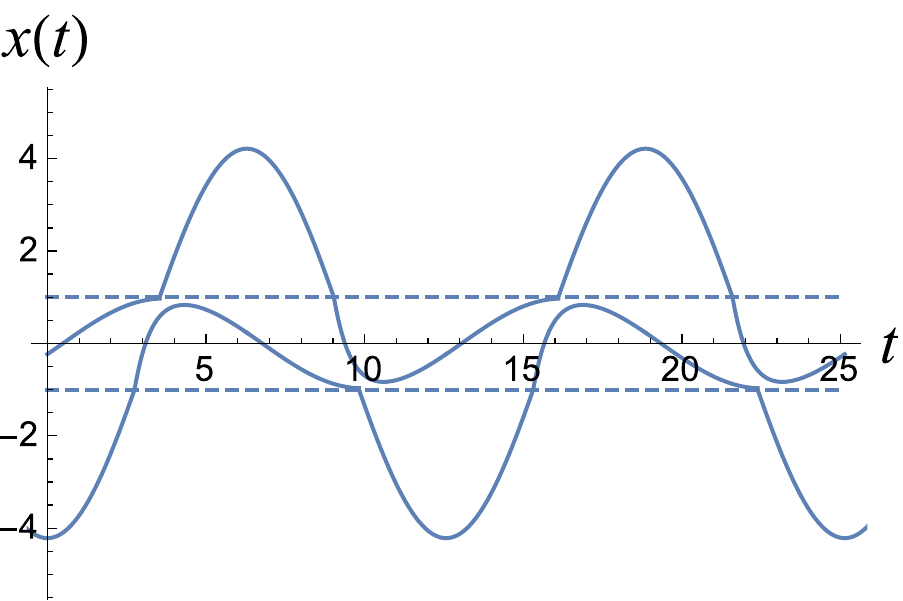}
\caption{Two skewed solutions to Eqs. (\ref{eq:eom1}) and (\ref{eq:forcelim1}). The value of $x_0 $ has been set equal to one.  }
\label{fig:coex1}
\end{center}
\end{figure}

Figure \ref{fig:coex1} shows two full periods of  two skewed solution to the equation of motion (\ref{eq:eom1}) with confining potential (\ref{eq:forcelim1}).
The displacement limits outside of which which the restoring force vanishes are indicated by dashed lines. Note the slope discontinuities when those limits are traversed. This is permissible in a system in which inertia is ignored. The trajectories consist of the two solutions (\ref{eq:inside}) and (\ref{eq:outside}), grafted together at the boundaries between their regimes of applicability. From the nature of the solution for $|x(t)| > x_0$, we see that, in the case of the solution skewed above the $x$ axis, the intervals in which the solution for $|x(t)| < x_0$  applies, as depicted in Fig. \ref{fig:coex1}, will satisfy
\begin{equation}
(2n+1) \pi / \omega + t_0 < t < (2n+3) \pi/ \omega -t_0 \, ,\label{eq:window}
\end{equation}
corresponding to time windows lying symmetrically within the interval between successive odd multiples of the period of the forcing term, $2 \pi/ \omega$. We can write for the form of a solution in such a time window
\begin{eqnarray}
\lefteqn{x(t)} \nonumber \\ &=&  -\frac{A}{\omega \sqrt{1+(k/\omega)^2}} \cos (\omega t+ \arctan(k/ \omega)) \nonumber \\ && + Ce^{-k\left( t -(2n+1) \pi/\omega - t_0 \right)} \label{eq:v6}
\end{eqnarray}
Here and henceforth, we have set the parameters $\gamma =1, x_0 = 1$. In order for the solution in (\ref{eq:v6}) to properly match the solution for $|x(t)|> x_0$ at the limits of the window, we require
\begin{eqnarray}
 x_0  & = & -\frac{A}{\omega \sqrt{1+(k/ \omega)^2}} \cos ( (2n+1) \pi + \omega t_0 \nonumber \\ && + \arctan(k/ \omega)) + C   \label{eq:v7} \\
x_0 & = & -\frac{A}{\omega \sqrt{1+(k/ \omega)^2}} \cos((2n+3) \pi - \omega t_0 \nonumber \\ && + \arctan(k/\omega)) + Ce^{-2 \pi k/ \omega + 2kt_0}  \label{eq:v8}
\end{eqnarray}
We can eliminate the constant $C$ between these two equations, and we are left with the equation for $t_0$
\begin{eqnarray}
\lefteqn{\frac{A}{\omega \sqrt{1+(k/ \omega)^2}} \bigg[ \cos(- \omega t_0 + \arctan (k/ \omega))} \nonumber \\ &&  - e^{- 2 \pi k/ \omega + 2kt_0} \cos( \omega t_0 + \arctan(k/ \omega)) \bigg]  \nonumber \\ && +x_0\left( e^{- 2 \pi k/ \omega +2 k t_0} -1 \right) =0 \label{eq:v9}
\end{eqnarray}

\begin{figure}[htbp]
\begin{center}
\includegraphics[width=3in]{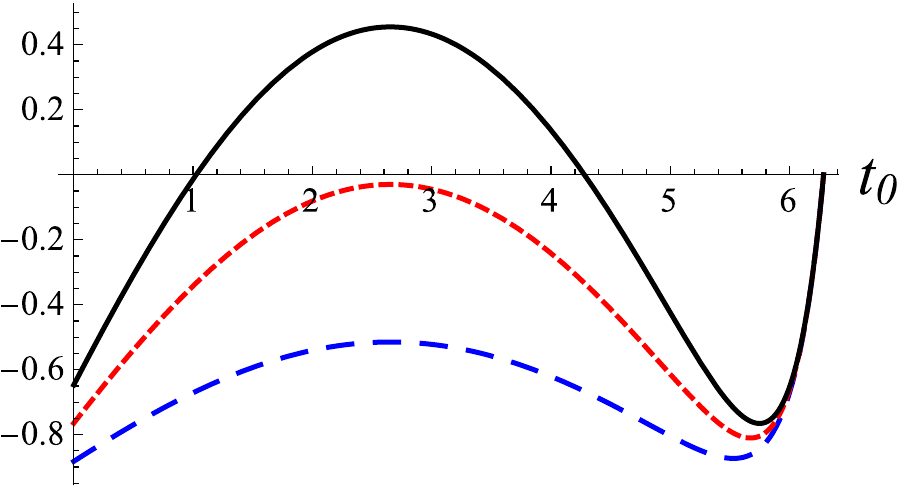}
\caption{The left hand side of (\ref{eq:v9}) as a function of $t_0$, with $x_0=1$, $k=2$ and $\omega=0.5$. The three curves correspond to the following three values of the drive amplitude: $A=1$ (blue long-dashed curve), $A=2$ (red dashed curve), $A=3$ (black solid curve).  }
\label{fig:t0plot}
\end{center}
\end{figure}

The left hand side of (\ref{eq:v9}) as function of $t_0$ is displayed in Fig. \ref{fig:t0plot}, with the variables $x_0$, $k$ and $\omega$ set so as to correspond to their values in the phase diagram in Fig. \ref{fig:phasediagrampic2}.
As shown in the figure, smaller amplitudes correspond to lower curves.
Note that there is always a solution
$\omega t_0 = \pi $, which says that the window in (\ref{eq:window}) has shrunk to zero. This is reasonable since it is always possible, by setting initial conditions appropriately, to construct a solution in which $x(t)$ lies completely in the regime of vanishing restoring force. When $t_0 = \pi/ \omega$  is the \textit{only} solution to (\ref{eq:v9}), then solutions to the equations of motion (\ref{eq:eom1}) with $F(x)$ given by (\ref{eq:forcelim1}) either lie completely in the regime $|x| < x_0 $ (where, in this example, $x_0=1$) or entirely outside of it. At a threshold value of the amplitude, two more apparent solutions arise, as for the solid curve in Fig. \ref{fig:t0plot}.

The next step is to determine whether the new intersections correspond to solutions of the equation of motion, and, if so, whether those solutions are dynamically stable. As it turns out, one of the two new solutions of (\ref{eq:v9})---the solution corresponding to the larger value of $t_0 $---does indeed correspond to a legitimate response. Furthermore, the response is dynamically stable. Appendix \ref{app:stability} describes the analysis.

\begin{figure}[htbp]
\begin{center}
\includegraphics[width=3in]{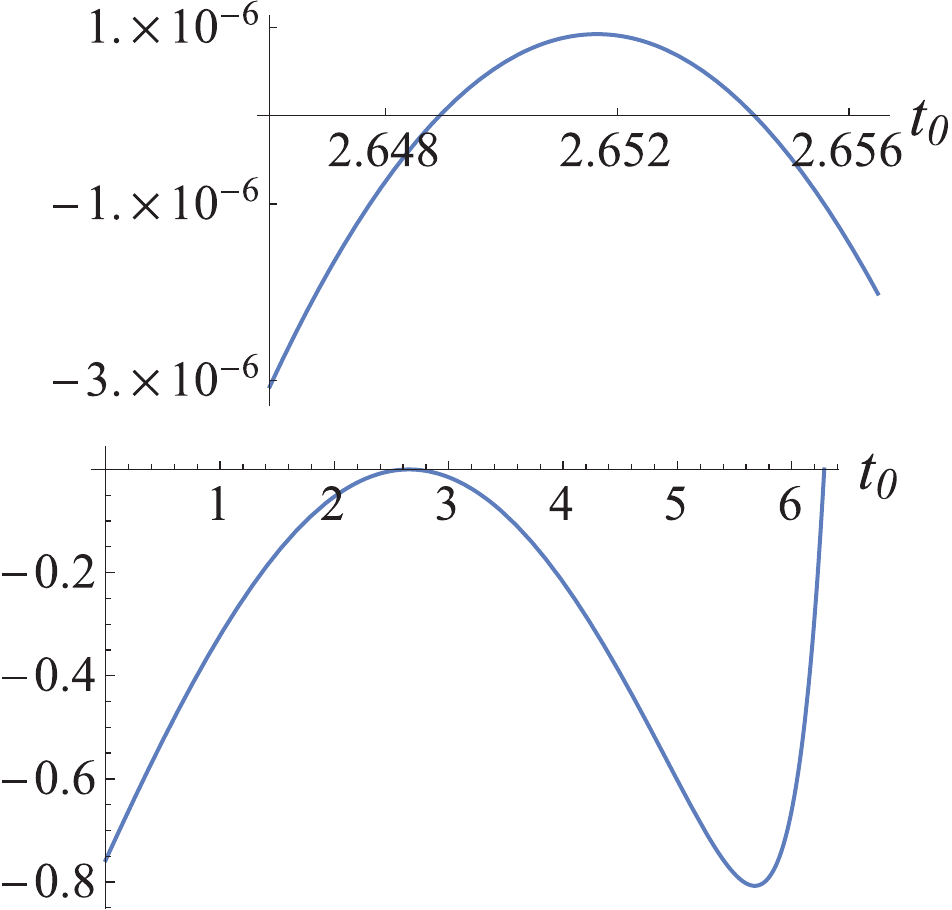}
\caption{The left hand side of Eq. (\ref{eq:v9}) when $k=2$, $\omega=0.5$, $\gamma =1$, $x_0=1$ and the drive amplitude $A$ is equal to $\sqrt{k^2+ \omega^2}$. The lower portion of the figure shows the expression for the entire range of $t_0$, from 0 to $\pi/\omega$. The upper portion magnifies the portion of the graph in which the left hand side of (\ref{eq:v9}) passes through zero.}
\label{fig:transtest1}
\end{center}
\end{figure}

\begin{figure}[htbp]
\begin{center}
\includegraphics[width=3in]{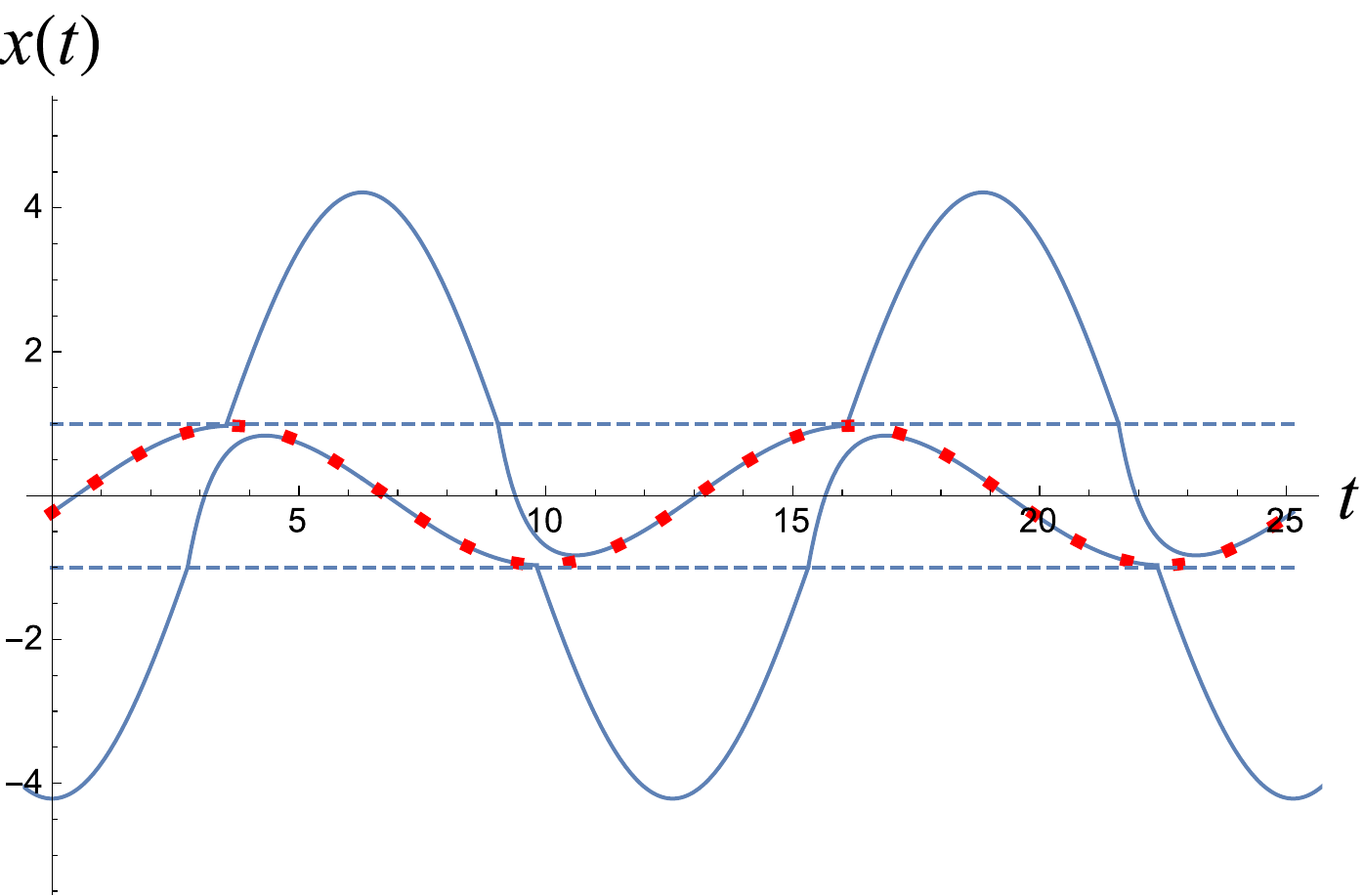}
\caption{The three stable solutions to the equation of motion (\ref{eq:eom1}) with restoring force given by (\ref{eq:forcelim1}). The parameters are the same as in Fig. \ref{fig:transtest1}. The two skewed solutions are shown as solid curves and the dots indicate the symmetric solution, which just fits in the Hooke's Law region for the restoring force. The boundaries of that region are shown as dashed lines. }
\label{fig:coexistence4}
\end{center}
\end{figure}

Another interesting aspect of the dynamical behavior is the existence of an exceedingly small range of amplitudes, $A$, in which a  dynamically stable, skewed solution coexists with a symmetric solution lying entirely inside the Hooke's law region. As an example of coexistence Figure \ref{fig:transtest1}
shows the left hand side of Eq. (\ref{eq:v9}) when the parameters are adjusted so that there is a solution to the equation of motion that just fits in the region $-x_0 < x < x_0$. Even though there is such a trajectory, an additional skewed and dynamically stable solution to the equation of motion (\ref{eq:eom1}) with $F(x)$ given by (\ref{eq:forcelim1}) also exists (actually a pair, see below), corresponding to the solution to (\ref{eq:v9}) in which $t_0 \approx 2.654$. The region of coexistence is spanned as the drive amplitude $A$ changes by about one part in $10^6$. Figure \ref{fig:coexistence4} shows the three stable solutions.
Not shown in the figure are the two unstable skewed solutions to the equation of motion that lie between the symmetric solution and the stable skewed solutions.

Given a solution, $x(t)$ of the equation of motion (\ref{eq:eom1}) with the kind of restoring force considered here, i.e. a restoring force with the property $F(-x) = -F(x)$, it is straightforward to show that the equation of motion is also satisfied by $-x(t +  \pi/\omega)$. From this we can infer from one skewed solution another one, skewed in the opposite direction.  This feature was used to construct the pair of skewed solutions in Fig.\ref{fig:coexistence4}. Thus, we have confirmed the existence of two distinct stable, skewed solutions to the equation of motion for sufficiently high values of the driving amplitude. See Appendix \ref{app:pairs}.

\section{Loss and storage compliances} \label{sec:compliances}

To further quantify the transition, we introduce a generalization of the way in which the response of a  linear viscoelastic system is characterized in terms of dissipative, or loss, compliances and storage, or elastic, compliances. Given the notation we utilize here, the stress exerted on the system is $ A \sin (\omega t)$. The response of the system, if linear, would be of the form
\begin{eqnarray}
x(t) &=& x_{\rm diss}(t) + x_{\rm el}(t) \nonumber \\
& = & -X_d \cos( \omega t) + X_e \sin( \omega t) \nonumber \\
& = & -j_2 A \cos( \omega t) + j_1 A \sin(\omega t) \, ,\label{eq:ds1}
\end{eqnarray}
where $j_1$ is the \textit{ storage compliance} and $j_2$ is the \textit{ loss compliance} \cite{Findley}, and the subscripts on the first line of (\ref{eq:ds1}) stand for ``dissipative'' and ``elastic.''

In the case of nonlinear response, it is possible to generalize the above decomposition by extracting the contribution to the response that gives rise to dissipation. We do this by taking the integral of the steady state solution over a period,
\begin{equation}
Y_{d }= \frac{\omega}{\pi}\int_0^{2 \pi/\omega} x(t) \cos(\omega t) d t \, ,\label{eq:ds2}
\end{equation}
in which case we can write
\begin{eqnarray}
x(t) &=& Y_d \cos(\omega t) + (x(t) - Y_d \cos(\omega t) ) \nonumber \\
& \equiv & x_{\rm diss}(t) + x_{\rm el}(t) \label{eq:ds3}
\end{eqnarray}
The entirety of the dissipative response is contained in $x_{\rm diss}(t)$ as defined above, in that
\begin{equation}
\int_0^{2 \pi/ \omega} \sin(\omega t) \frac{dx_{\rm el}(t)}{dt} dt =0 \label{eq:ds4}
\end{equation}
This means that the quantity $-Y_d/\sqrt{2}A$ (see below) plays the role of the dissipative compliance, while there is not necessarily any quantity that can be unambiguously associated with the storage compliance.

\begin{figure}[htbp]
\begin{center}
\includegraphics[width=3in]{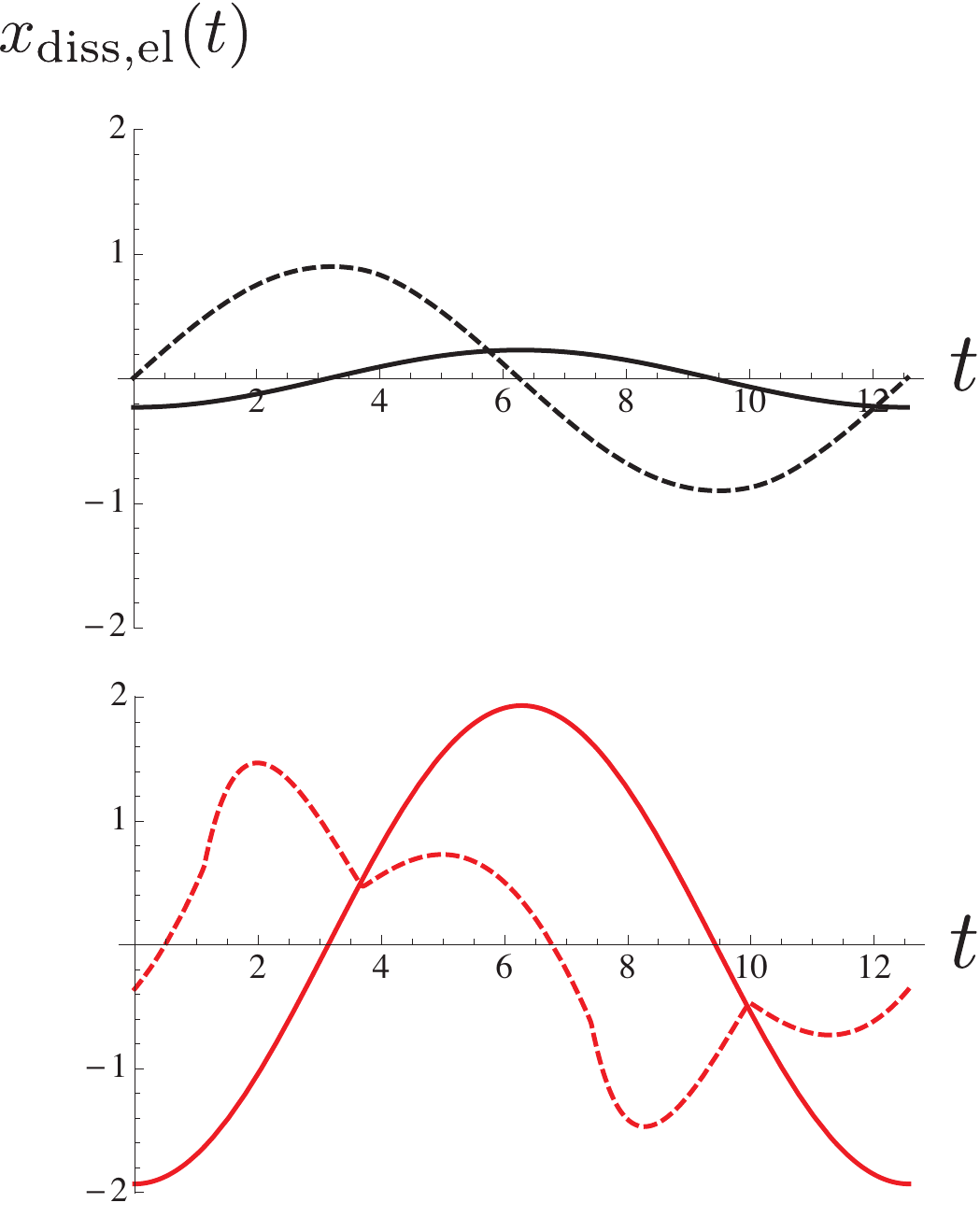}
\caption{The two solutions for $x(t)$ shown in Fig. \ref{fig:1and1a}, broken into dissipative contributions (solid curves) and storage contributions (dashed curves); see Eq. (\ref{eq:ds3}). The top plot shows the solution in Region 1, in which the response is dominantly elastic and the bottom plot shows the solution in Region 1a in which the response is more dissipative. Recall that the inverse of the drive amplitude is 0.53 in the case of the top plot and 0.51 for the bottom plot,
and $F_0 = 0.6$.}
\label{fig:belowandabove}
\end{center}
\end{figure}

We now return to our "standard model" shown schematically in Fig. \ref{fig:enforce}. Figure \ref{fig:belowandabove} displays the two solutions for $x(t)$ shown in Fig. \ref{fig:1and1a}, divided into dissipative and elastic components. As is clear from the lower plot in the figure, which refers to the response in Region 1a, the elastic response is not simply proportional to the driving force, $A \sin(\omega t)$.

However, based on the developments above, we can reduce the elastic and dissipative components of the response to two numbers. If
\begin{equation}
I_t^2 = \frac{\omega}{2 \pi}\int_0^{2 \pi/\omega} x(t)^2 dt \, ,\label{eq:ds5}
\end{equation}
then
\begin{eqnarray}
\frac{\omega}{2 \pi} \int_0^{2 \pi/ \omega} x_{\rm diss}(t)^2 dt &=& \frac{Y_d^2}{2} \nonumber \\ & \equiv & I_d^2 \label{eq:ds6} \\
\frac{\omega}{2 \pi} \int_0^{2 \pi/ \omega} x_{\rm el}(t)^2 dt & = & I_t^2 -I_d^2 \nonumber \\ & \equiv & I_s^2 \, ,\label{eq:ds7}
\end{eqnarray}
which means we can define an overall compliance as $I_t/A$, with a storage compliance, $j_1 = I_s/A$, and a loss compliance, $j_2 = I_d/A$.

As a final technicality, we need to address the case in which there are dynamically stable skewed solutions to the equation of motion. The measurements that motivated this study are made on a collection of driven oscillators; furthermore, thermal noise is substantial. As we will see such a system will equilibriate into an ensemble in which the two skewed responses, when they exist, are equally represented. This means that under such conditions, we should replace $x(t)$ in the equations defining the compliance by
\begin{equation}
\langle x(t) \rangle = (x(t) - x(t + \pi/ \omega))/2
\label{eq:skave}
\end{equation}

\begin{figure}[htbp]
\begin{center}
\includegraphics[width=3in]{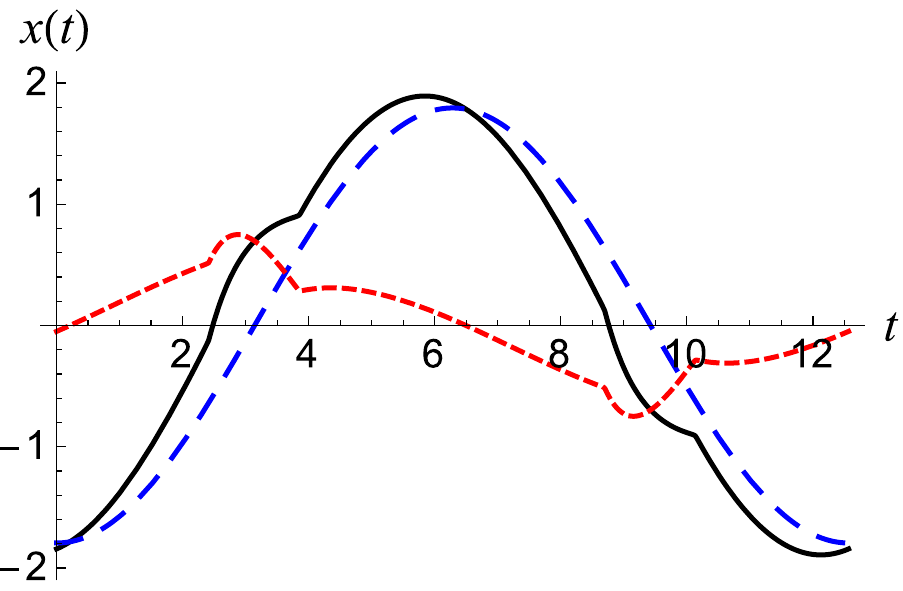}
\caption{The average of the  two stable skewed response curves. The solid black curve is the full response. The long dashed blue curve is the dissipative contribution, going as $\cos( \omega t)$, which we term $x_{\rm diss}(t)$. The dashed red curve is what remains when the dissipative component is subtracted from the full response. We term that response $x_{\rm el}(t)$ (See (\ref{eq:ds3}). Here, $k=2$, $\omega =0.5$, $F_0 =0$,  $\alpha =0.01$ and $A$, the drive amplitude, is equal to 1.9. }
\label{fig:responsecomposition2}
\end{center}
\end{figure}

Figure \ref{fig:responsecomposition2} displays an example of the case in which the response curve is skewed and the average in (\ref{eq:skave}) has been performed  The full response is shown as well as the two components that comprise it.

\section{On the nature of the dynamical phase transition} \label{sec:transnature}

\begin{figure}[htbp]
\begin{center}
\includegraphics[width=3in]{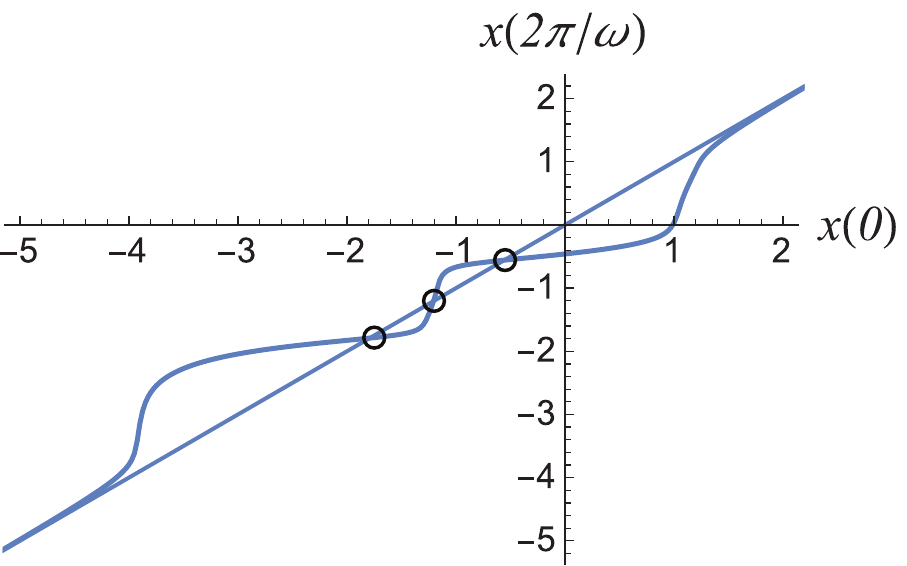}
\caption{Example of the map connecting $x(0)$ with $x(2 \pi/\omega)$, the quantity $x(t)$ satisfying Eq. (\ref{eq:eom1}).  Also shown is the line $x( 2 \pi/\omega) = x(0)$. In this case, the drive amplitude, $A$ is 2.75, the frequency, $\omega$, of the drive is 2, $k$ is 2, the parameter $\alpha$ is 0.15 and the asymptotic force, $F_0$, is equal to zero. As always, $x_0=1$ and $\gamma =1$. The small open circles denote the intersections between the map and the $45^{\circ}$ line corresponding to steady state solutions of the equation of motion.   }
\label{fig:recur1}
\end{center}
\end{figure}

Our focus has been the steady state behavior of the dynamical system described by (\ref{eq:eom1})
When the crossover from a restoring force described by Hooke's Law to a constant restoring force $F_0$ is not abrupt, the methods of Sec. \ref{sec:analytical} cannot be applied. An alternate approach that proves powerful, useful and universal, utilizes a  standard map or recursion relation. This relation follows from the fact that the solution of a first order differential equation like (\ref{eq:eom1}) is determined entirely by a single initial condition. That is, if we know $x(0)$, then we can use the equation to forward integrate and determine $x(t)$ at all subsequent times. In particular, we can create a map that takes us from $x(0)$ to $x(2 \pi/\omega)$, the value of the dynamical variable one period later. A steady state solution to the equation of motion will have the property $x(2 \pi/\omega)= x(0)$---if we can discount the possibility of solutions with a period at a subharmonic of the driving force. Figure \ref{fig:recur1} displays such a recursion graph, calculated from Eq. (\ref{eq:eom1}) with a particular choice of parameter values.
Also shown in that figure is the $45^{\circ}$ line corresponding to $x(2 \pi/\omega) = x(0)$. The small open circles indicate intersection of that line with the recursion curve, corresponding to steady state solutions of the equation.

\begin{figure}[htbp]
\begin{center}
\includegraphics[width=2.5in]{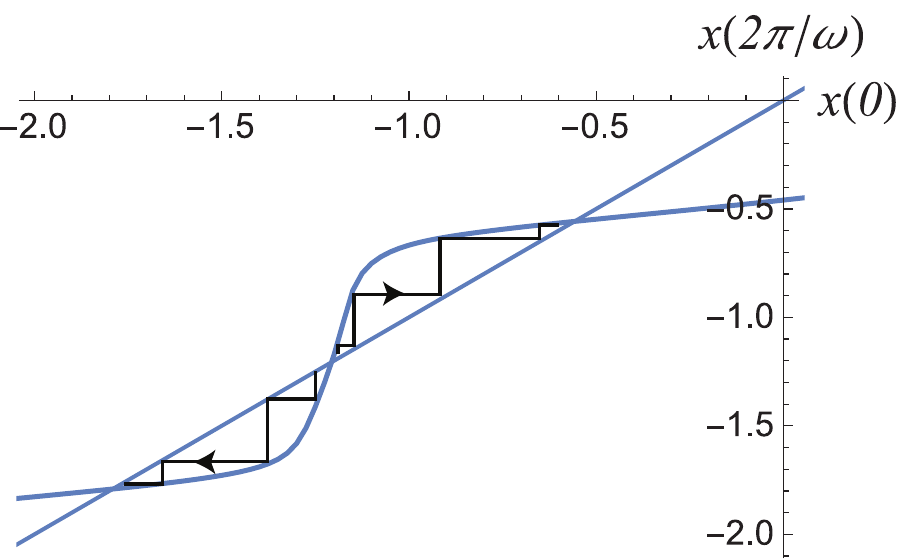}
\caption{The result of graphical iterations of the recursion relation embodied in the curve displayed in Fig. \ref{fig:recur1}. The broken curves impinge on the recursion curve vertically and on the $45^{\circ}$ line horizontally. }
\label{fig:stability1}
\end{center}
\end{figure}

We can assess the dynamical stability of the steady state solutions in the standard fashion by iterating the recursion relation \cite{GandH,Feigenbaum}. The graphical version of this process is displayed in Fig. \ref{fig:stability1}, in which we focus our attention on the section of the curve containing the intersections.
The stepwise curves between the map and the $45^{\circ}$ line are graphical renditions of the result of iteration of the recursion relation. As indicated by the arrowheads on those curves, repeated calculations of the quantity $x(t)$ at succeeding intervals of one period tend away from the central intersection and towards one of the flanking ones. We are led to the conclusion that the central solution for a steady state is dynamically unstable, while the two solutions that flank it are, by contrast, dynamically stable. Such a recursion relation yields the solutions one finds in Region 2 of the dynamical phase diagram as shown in  Fig. \ref{fig:phasediagrampic2}.

\begin{figure}[htbp]
\begin{center}
\includegraphics[width=3in]{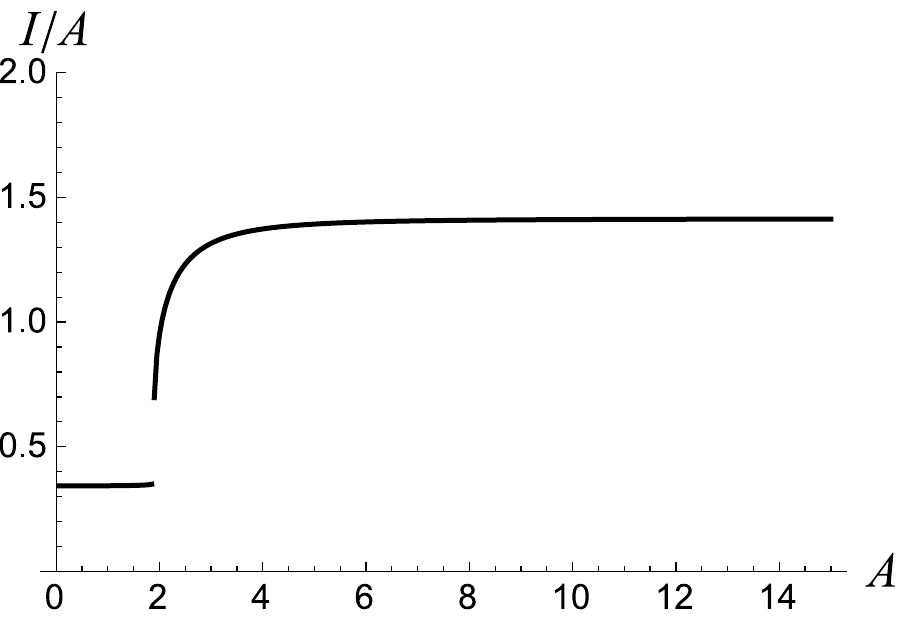}
\caption{The total compliance, $I_t/A$, plotted agains the drive amplitude, $A$. Here, $F_0$=0, $\omega=0.5$, $k=2$ and the parameter $\alpha = 0.01$. Here and in Fig. \ref{fig:phaset2} the compliances are as defined in Sec. \ref{sec:compliances}. }
\label{fig:phaset1}
\end{center}
\end{figure}

\begin{figure}[htbp]
\begin{center}
\includegraphics[width=3in]{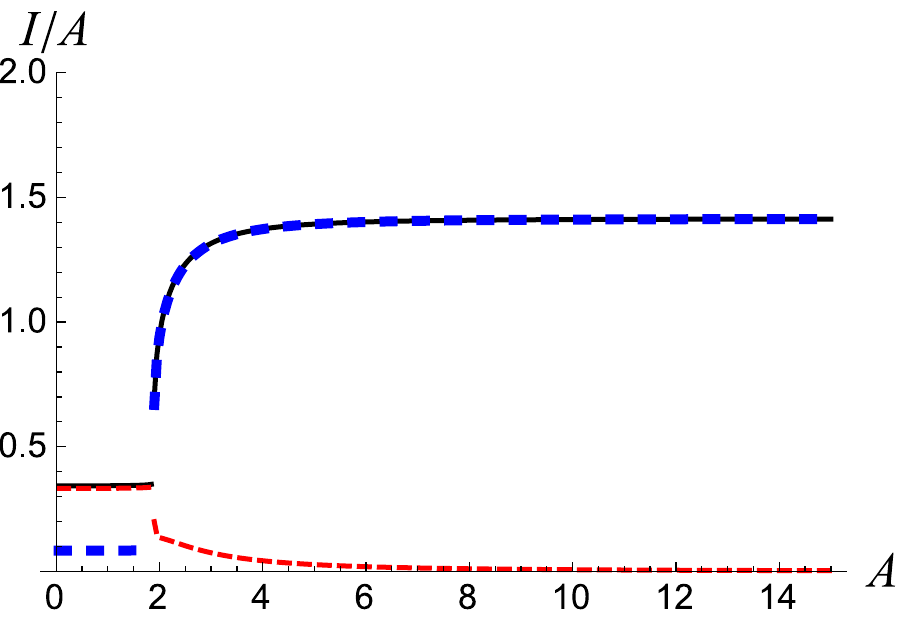}
\caption{The two contributions to the total compliance graphed in Fig. \ref{fig:phaset1}: the dissipative compliance $I_d/A$ (long dashed blue curve),  the storage compliance $I_s/A$ (dashed red curve) and the total compliance, (thin black curve).}
\label{fig:phaset2}
\end{center}
\end{figure}

Given a steady state solution for a particular $x(0)$, we can then compute $x(t)$ for the entire interval between $t=0$ and $t= 2 \pi/\omega$, and from this the response properties of that solution.  As an example, we can explore in more detail the transitions between the various regions shown in the phase diagram in Fig. \ref{fig:phasediagrampic2}. We start by looking at the portion of the phase diagram that is exactly on the vertical axis, i.e. for which the large-$x$ asymptote of the restoring force is $F_0 = 0$. Making use of the definitions in Sec. \ref{sec:compliances} for total, storage and loss compliances, we obtain results for the quantities $I_t/A$, $I_s/A$ and $I_d/A$. Figure \ref{fig:phaset1} shows the total compliance, $I_t/A$, plotted in terms of the amplitude, $A$, of the drive.
Note the discontinuity in the compliance, which can be taken as evidence for a first order transition. Figure \ref{fig:phaset2} shows the two contributions to the total compliance, obtained from $I_d/A$, and $I_s/A$.

 Close investigation reveals a very narrow regime around the point of discontinuity in which there is coexistence between the responses corresponding to the compliances to the right and left of the transition point in Figs. \ref{fig:phaset1} and \ref{fig:phaset2}.  The regime is sufficiently thin that it is undetectable given the resolution of the figure.

The discontinuities in the dynamic response graphed in Figs. \ref{fig:phaset1} and \ref{fig:phaset2} imply a dynamical transition with the characteristics of a first order thermodynamic phase transition, in that there are discontinuities in key quantities. However, a transition with discontinuities is not inevitable. If the change in the restoring force from harmonic to a constant (in the example being discussed: $F_0=0$) is sufficiently gentle, then the dynamical transformation becomes\textit{ continuous} in the thermodynamic sense. That is, it remains sharp, but physical properties, such as dynamical moduli, no longer exhibit discontinuities in their dependence on the drive amplitude.

\begin{figure}[htbp]
\begin{center}
\includegraphics[width=3in]{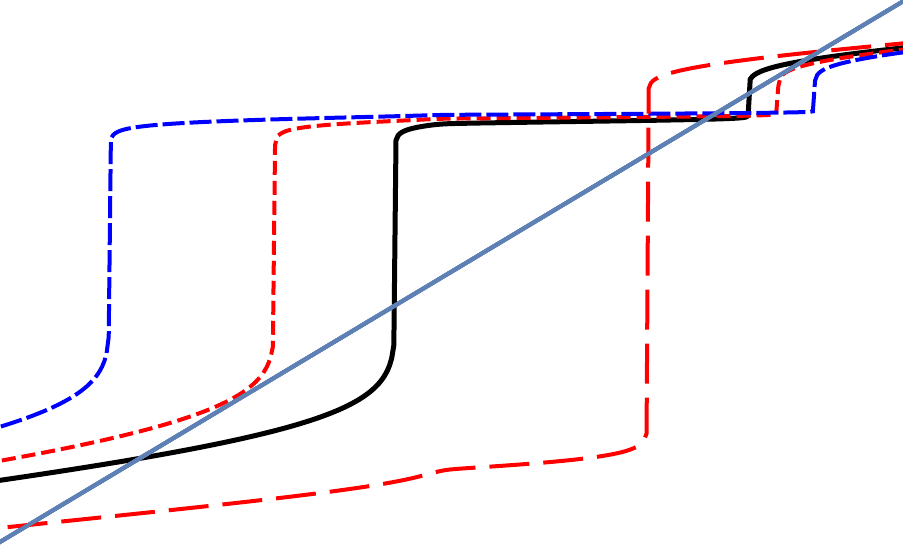}
\caption{Illustrating a first order transition between Region 1 and Region 2. Referring to Appendix \ref{app:restore}, the parameters in the restoring force and the equation of motion are as follows: $\omega=2$, $k=2$, $x_0=1$, $\alpha=0.01$ and $F_0=0$. The longer dashed blue curve and the shorter dashed red curve correspond to Region 1 in the phase diagram in Fig. \ref{fig:phasediagrampic2}. The longest dashed red curve corresponds to Region 2, and the solid black curve to a coexistence region between Regions 1 and 2. Such a region exists but is exceedingly narrow for the set of variables used to generate Fig. \ref{fig:phasediagrampic2}.  }
\label{fig:fograph}
\end{center}
\end{figure}

\begin{figure}[htbp]
\begin{center}
\includegraphics[width=3in]{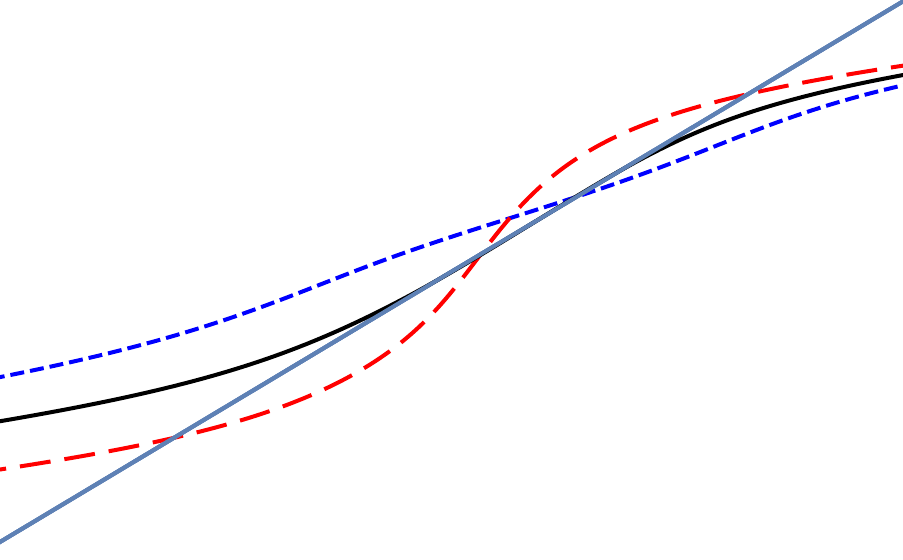}
\caption{Illustrating a continuous transition between Regions 1 and 2 in Fig. \ref{fig:phasediagrampic2} when $\alpha$ is increased from 0.01 to 0.15 so as to stretch out the transition region between a Hooke's Law restoring force and a vanishing restoring force, i.e. $F_0=0$. Otherwise the parameters utilized are the same as in Fig. \ref{fig:criticalgraph}. See Appendix A. The dashed blue curve corresponds to Region 1 and the long dashed red curve to Region 2. The solid black curve corresponds to the onset of the transition between the two regions. }
\label{fig:criticalgraph}
\end{center}
\end{figure}

The change in the character of the transition as the properties of the restoring force are altered is illustrated in Figs \ref{fig:fograph} and \ref{fig:criticalgraph}, which display plots highlighting the intersections of recursion curves with the $45^{\circ}$ line. In particular, the figures show how the points of intersection change with drive amplitude. The plot in Fig. \ref{fig:fograph} contains four recursion curves, corresponding to four different drive amplitudes. In this case there is a first order dynamical phase transition. The parameters are---with the exception of $\omega$---the same as in the phase diagram in Fig.  \ref{fig:phasediagrampic2} with $F_0=0$. The drive frequency has been set as $\omega = 2$ in order to make certain features of the curve more visible and to expand the coexistence region between Regions 1 and 2. The longer dashed blue curve intersects the $45^{\circ}$ line once, corresponding to a single symmetric and stable dynamical steady state. This curve is characteristic of low drive amplitude and (more nearly) elastic response. The (short-) dashed red curve corresponds to the drive amplitude at which one sees the onset of four more steady state solutions to the equation of motion, all of them skewed, two stable and two unstable. In the regime represented by the solid black curve, those additional solutions are clearly visible, and three dynamically stable steady state solutions to the equation of motion coexist -- one symmetric and two skewed. The long-dashed red curve corresponds to the high amplitude regime, Region 2 in the phase diagram, in which there are two stable skewed steady state solutions and one unstable symmetric solution to the equation of motion. The restoring force is as given in Appendix \ref{app:restore}, with parameter $\alpha =0.01$, the same value as was used to generate the phase diagram in Fig. \ref{fig:phasediagrampic2}.

Figure \ref{fig:criticalgraph}  displays the effect of smoothing out the transition region in $F(x)$. Here, the parameter $\alpha$ has been set equal to 0.15. In this case, the transition from a single, stable and symmetric, steady state solution to three steady state solutions---two skewed and stable and one symmetric and unstable---occurs continuously with no coexistence region and no discontinuities in dynamical response. The drive frequency, $\omega$, has again been set equal to 2, while the spring constant $k$ in the Hooke's Law region is maintained at 2. The short-dashed blue curve corresponds to Region 1, the long-dashed red curve is characteristic of Region 2, and the solid black curve illustrates the onset of the transition between one symmetric, stable solution to the equation of motion and three solutions, two skewed and stable, one symmetric and unstable. No region of coexistence separates the two.

\begin{figure}[htbp]
\begin{center}
\includegraphics[width=3in]{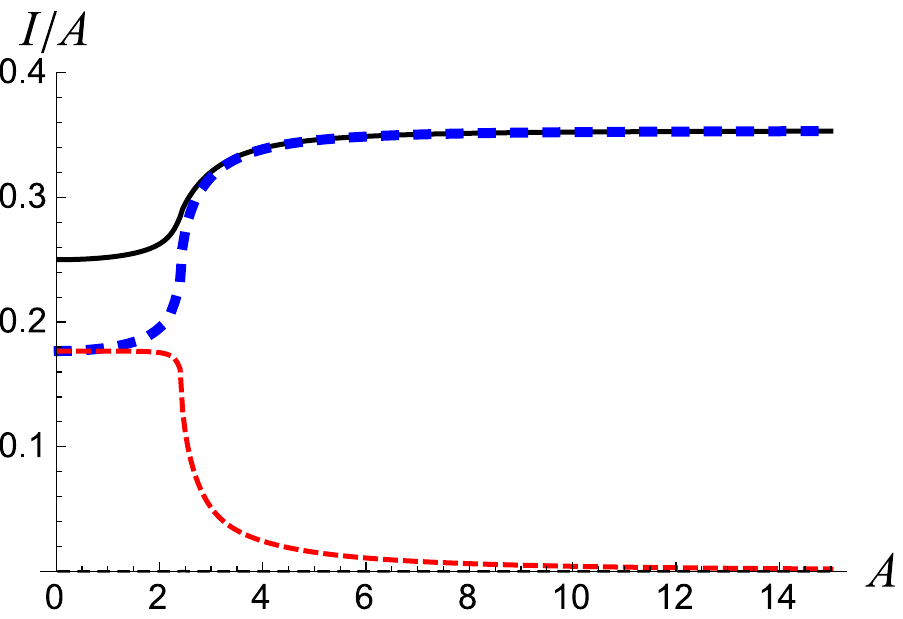}
\caption{In the case of parameter values leading to the plot in Fig. \ref{fig:criticalgraph}, the  total compliance, $I_t/A$, (solid black curve) and the three contributions to that compliance: the dissipative compliance $I_d/A$ (long dashed blue curve) and  the storage compliance $I_s/A$ (dashed red curve). Here, the transition is continuous. Also shown on the graph is the value of the amplitude, $A$ at which the transition between Section 1 and Section 2 in the phase diagram occurs. For definitions of the compliances see Sec. \ref{sec:compliances}.}
\label{fig:continuous}
\end{center}
\end{figure}

Figure \ref{fig:continuous} illustrates the compliances associated with Fig \ref{fig:criticalgraph}, quantified in terms of those defined in Sec. \ref{sec:compliances}.
Because of the particular relative values of $\omega$ and $k$, the compliance in Region 1 of the phase diagram divides equally into viscous and elastic, while in Region 2 viscous response increasingly dominates.

\begin{figure}[htbp]
\begin{center}
\includegraphics[width=3in]{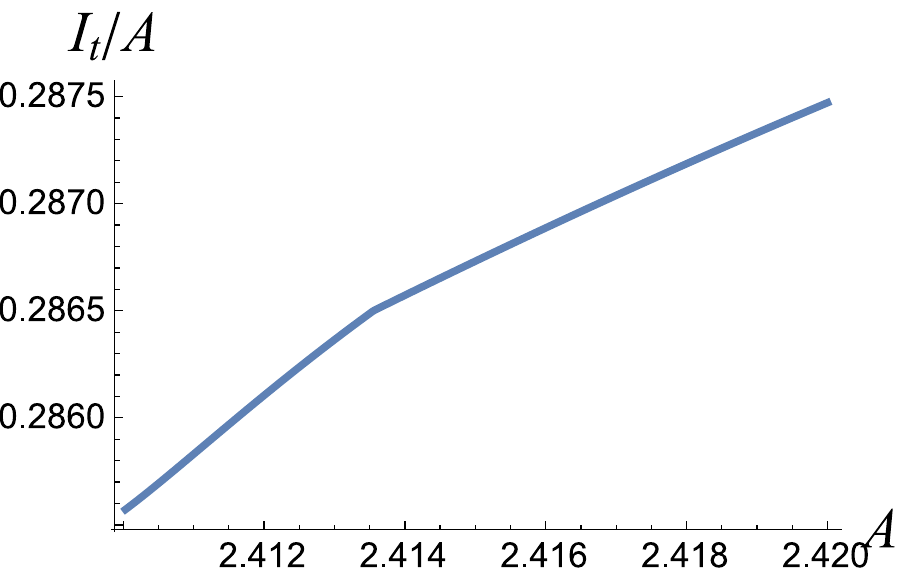}
\caption{The total compliance, $I_t/A$, in the immediate vicinity of the continuous transition from Region 1 to Region 3 in the dynamical phase diagram. }
\label{fig:detail}
\end{center}
\end{figure}

The compliance curves are continuous and without evident features marking the transition from Region 1 to Region 2. Very close inspection reveals singularities in the form of mild slope discontinuities in $I_t/A$, $I_s/A$ and $I_d/A$, which are unlikely to be detectable in any experimental realization of this system. Figure \ref{fig:detail} shows the total compliance, $I_t/A$, in the immediate vicinity of the transition.

The recursion curves bear a close relationship to the free energy extremum curves that arise in a simple order parameter-based model of a tricritical system. To be specific, the intersection of the recursion curves with the $45^{\circ}$ line correspond with the intersection of the curves for $d \mathcal{F}(\psi)/d \psi$ with the abscissa, as shown in Figs. \ref{fig:tc1a} and \ref{fig:tc2a} in Appendix \ref{app:tricrit}, which reviews the classical order parameter based model of a tricritical point.

\section{On the effects of inertia} \label{sec:inertia}

\begin{figure}[htbp]
\begin{center}
\includegraphics[width=3in]{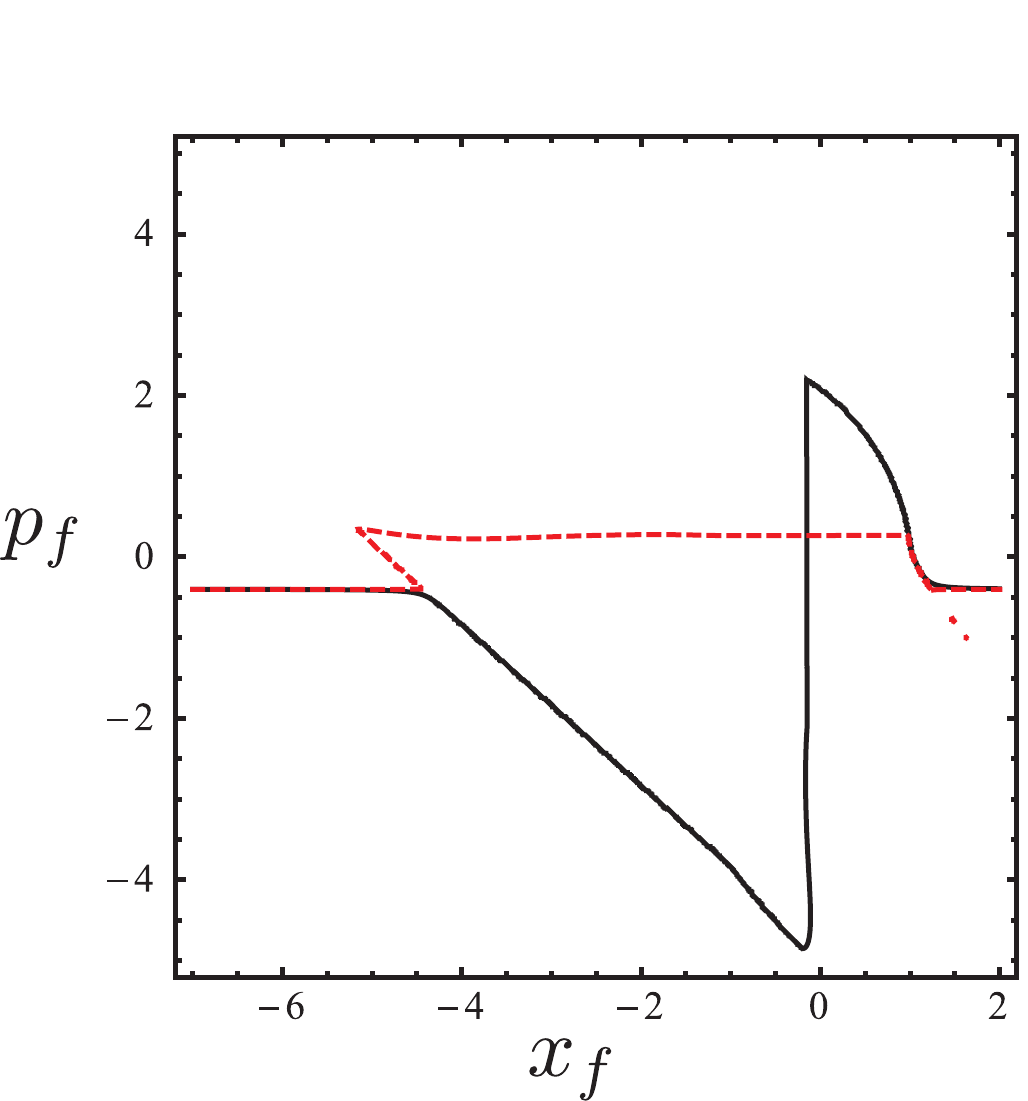}
\caption{A curve corresponding to (\ref{eq:in6}) (solid curve) and (\ref{eq:in7}) (dashed curve). In this case, we have taken $k=2$, $\gamma=1$, $\alpha=0.01$ in the expression for the restoring force (see Appendix \ref{app:restore}) and the force outside the Hookean region, $F_0=0$. The effective mass is $m =1$, and the drive amplitude, $A=1.82$.}
\label{fig:infig1}
\end{center}
\end{figure}

\begin{figure}[htbp]
\begin{center}
\includegraphics[width=3in]{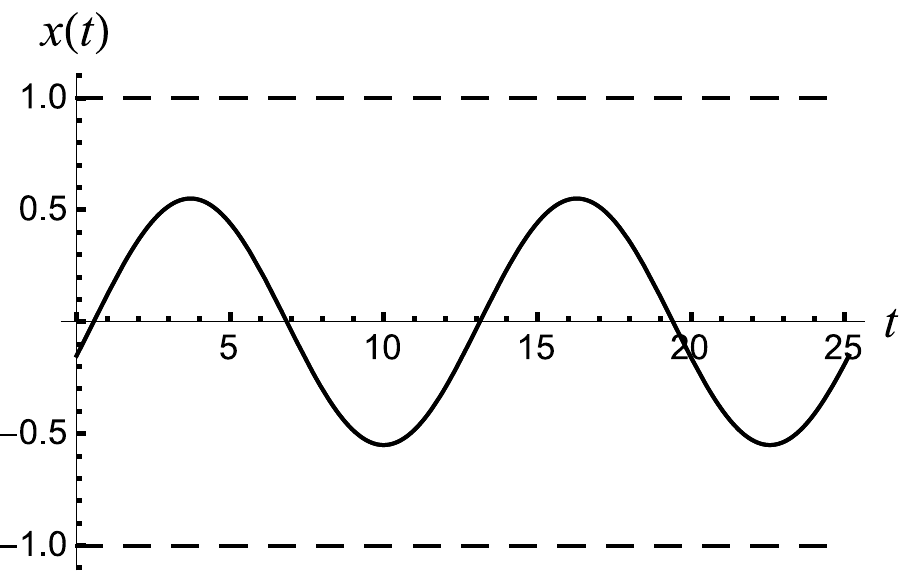}
\caption{The solutions, $x(t)$, corresponding to the  point of intersection in Fig. \ref{fig:infig1}. The response is entirely inside the regime in which the restoring force is Hookean.  }
\label{fig:infig2}
\end{center}
\end{figure}

So far we have ignored any effects of inertia on the response of the system. In light of the unexpected nature of the over-damped response and the typically significant effects of inertia,
including resonant behavior, it is worthwhile to ask whether the dynamical transitions observed in the absence of inertia survive its introduction into the equations of motion. To this end, we have performed preliminary studies of an extension of (\ref{eq:eom1}) that incorporates an inertial term.  The new equation of motion is
\begin{equation}
m \frac{d^2x(t)}{dt^2} + \gamma \frac{dx(t)}{dt} =F(x(t)) + A \sin(\omega t)\, , \label{eq:in1}
\end{equation}
which can be rewritten as
\begin{eqnarray}
\frac{dx(t)}{dt} & = & \frac{p(t)}{m} \label{eq:in2} \\
\frac{dp(t)}{dt} & = & - \gamma x(t) +F(x(t)) + A \sin( \omega t) \label{eq:in3}
\end{eqnarray}
The search for steady state solutions to the above equation of motion can be formulated in terms of a pair of recursion relations. That is, on the basis of Eqs. (\ref{eq:in2}) and (\ref{eq:in3}) we can construct a map from $x_0 = x(0)$ and $p_0=p(0)$ to $x_1=x( 2 \pi/\omega)$ and $p_1=p( 2 \pi/\omega)$. The map takes the form
\begin{eqnarray}
x_1 & = & X(x_0,p_0) \label{eq:in4} \\
p_1 & = & P(x_0,p_0) \label{eq:in5}
\end{eqnarray}
Steady state solutions to the equation of motion are $x_f$ and $p_f$, where
\begin{eqnarray}
x_f & = & X(x_f,p_f) \label{eq:in6} \\
p_f & = & P(x_f,p_f) \label{eq:in7}
\end{eqnarray}
Taken separately each of the relations above generates a curve in the ($x_f$, $p_f$) plane. Figure \ref{fig:infig1} displays two such curves.
The  point at which the two curves intersect corresponds to a simultaneous solution of (\ref{eq:in6}) and (\ref{eq:in7}) and thus a steady state solution of (\ref{eq:in2}) and (\ref{eq:in3}). The solution $x(t)$ arising from that intersection, which corresponds to a relatively weak driving force, is displayed in Fig. \ref{fig:infig2}.
Also shown in that figure are the nominal limits of the harmonic restoring force, at $x=\pm1$.

\begin{figure}[htbp]
\begin{center}
\includegraphics[width=3in]{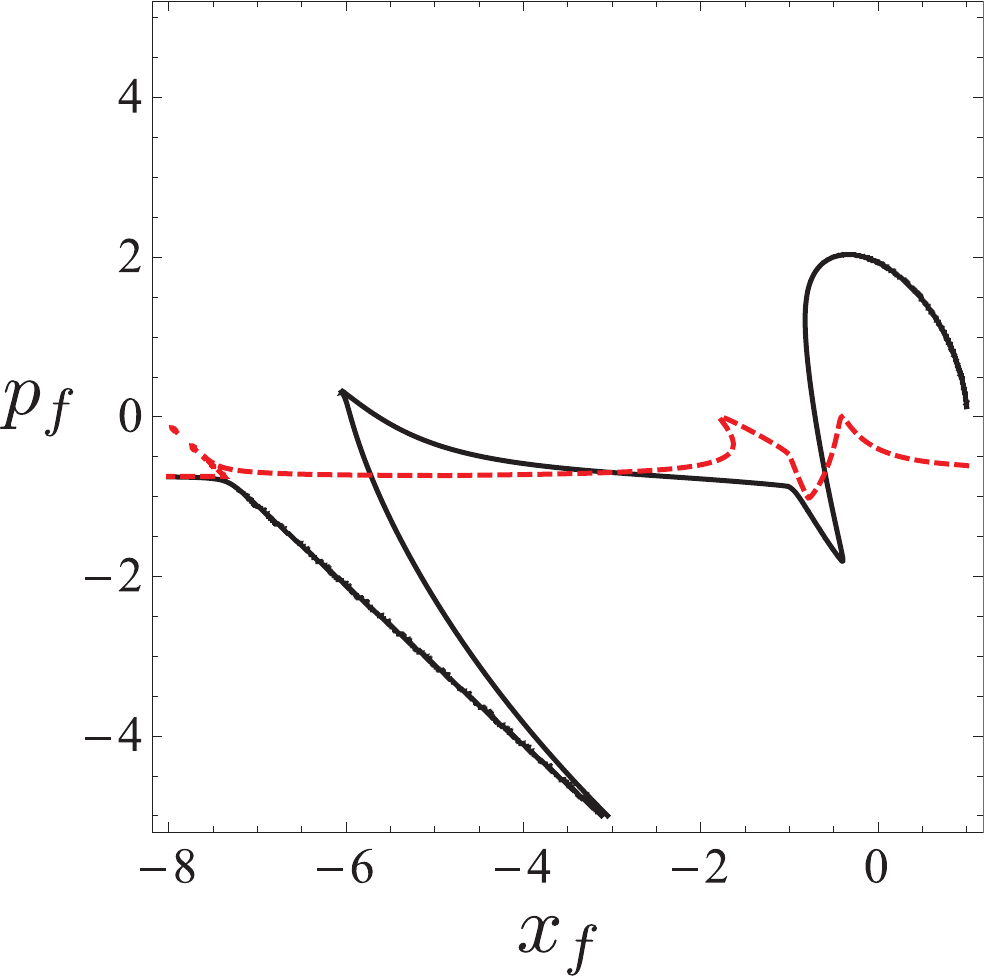}
\caption{Another pair of curves corresponding to (\ref{eq:in6}) (solid curve) and (\ref{eq:in7}) (dashed curve). In this case the parameters in the equation of motion (\ref{eq:in1}) are the same as in Fig. \ref{fig:infig1}, except for the drive amplitude, $A$, which is now 1.87. Note the presence of three intersections of the two curves.  }
\label{fig:infig1pp}
\end{center}
\end{figure}

\begin{figure}[htbp]
\begin{center}
\includegraphics[width=3in]{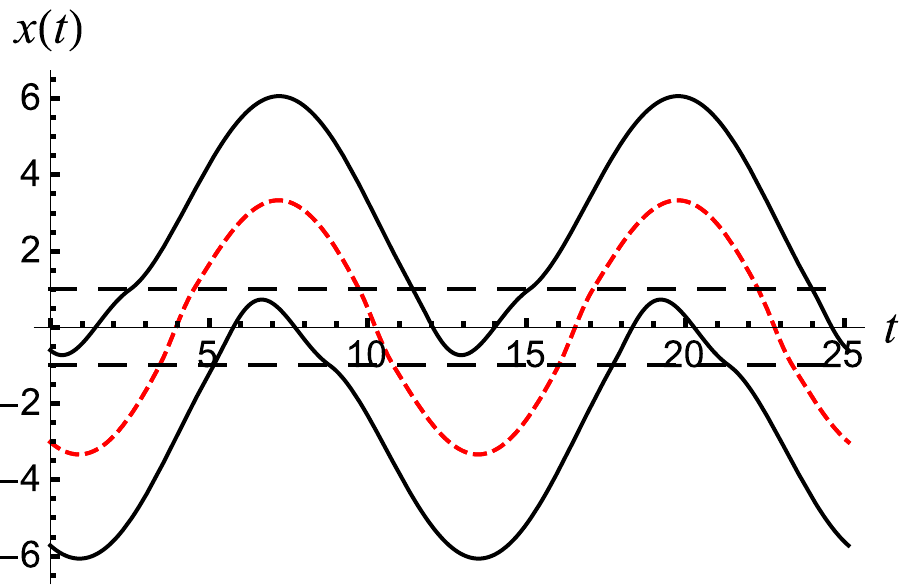}
\caption{The three solutions corresponding to the intersections in Fig. \ref{fig:infig1pp}. The two solid curves are the two stable, skewed solutions to the equation of motion (\ref{eq:in1}), while the dashed curve is a symmetric and unstable solution to that equation. The horizontal lines indicate the limit of the region in which the restoring force is Hookean. }.
\label{fig:infig2pp}
\end{center}
\end{figure}

By contrast, when the amplitude of the drive increases, more than one solution to the fixed point equations (\ref{eq:in6}) and (\ref{eq:in7}) appear, as shown in Fig. \ref{fig:infig1pp}.
The curves for $x(t)$ corresponding to the three intersections are shown in Fig. \ref{fig:infig2pp}. The relationship between the two skewed solutions is the same as the relationship between ``mirror image'' solutions established for solutions to (\ref{eq:eom1}) in Appendix \ref{app:pairs}. In fact, the argument in  Appendix B is easily extended to incorporate the inertial term in the equation of motion (\ref{eq:in1}).

The question of the stability of the solutions of (\ref{eq:in6}) and (\ref{eq:in7}) can be addressed in a straightforward manner. Imagine that $x_F$ and $p_F$ are such simultaneous solutions. Then, if $x_0=x_F + \Delta x$ and $p_0=p_F + \Delta p$,
\begin{eqnarray}
x_1 & = &x_F+ \Delta x^{\prime} \nonumber \\ & = & X(x_F+ \Delta x, p_F + \Delta p) \nonumber \\ & = & x_F + X_x \Delta x + X_p \Delta p \label{eq:ex1} \\  p_1 & = & p_F + \Delta p^{\prime} \nonumber \\& = & p_F + P_x \Delta x + P_p \Delta p \label{eq:ex2}
\end{eqnarray}
or
\begin{equation}
\left( \begin{array}{l} \Delta x^{\prime} \\ \Delta p^{\prime} \end{array} \right) = \left( \begin{array}{ll} X_x & X_p \\ P_x & P_p \end{array} \right) \left(\begin{array}{ll} \Delta x \\ \Delta p \end{array} \right) \label{eq:ex3}
\end{equation}
from which we infer that the stability of the solutions is controlled by the eigenvalues of the matrix on the right hand side of (\ref{eq:ex3}). If both have absolute value less than 1, the fixed point at $x_F$, $y_F$ is stable; otherwise it is unstable. It is possible to determine the elements of the matrix numerically by exploring the recursion relations in the vicinity of the fixed point. Using this analysis, we find that the single intersection shown in Fig. \ref{fig:infig1} is stable and that the intersections on the far right and far left of Fig. \ref{fig:infig1pp} also represent dynamically stable solutions to the simultaneous equations (\ref{eq:in6}) and (\ref{eq:in7}) -- while the central intersection, which gives rise to the steady state solution shown dashed in Fig. \ref{fig:infig2pp}, corresponds to a dynamically unstable solution to the equation of motion (\ref{eq:in1}).

An alternate approach to the stability analysis of the steady state solutions to (\ref{eq:in1}) is described in Appendix \ref{app:instability}.

The above results strongly indicate the the types of dynamical transitions that we find when inertia is
neglected are also to be expected in a system that manifests inertia.  A more comprehensive exploration of the nature of viscoelastic behavior in such systems, including the range of dynamical transitions that one may encounter, remains to be undertaken.

\section{Noise} \label{sec:noise}

\subsection{Langevin equation approach} \label{subset:langevin}

As noted in the Introduction our original motivation was a series of experiments on the mechanical properties of a driven natively folded protein in solution at room temperature. Clearly this is a noisy environment, and it behooves us to examine which features we have uncovered survive the introduction of fluctuations, thermal and otherwise. In this section we briefly and qualitatively consider some of the modifications that arise when noise is introduced into the over-damped equation of motion, Eq. (\ref{eq:eom1}).
For a specific example, we take the nonlinear restoring force to be the one defined in Appendix A, with $F_0=0, \alpha=0.01$. The restoring force is linear in displacement for $|x|$ less then about unity, decaying to zero restoring force outside that interval, as sketched in Fig. \ref{fig:enforce} with $F_0 = 0$.

With noise the dynamical equation is
\begin{equation}
\frac{dx(t)}{dt} = \frac{1}{\gamma} [-V^{\prime}(x(t)) + A \sin ( \omega t)] + \eta(t) \label{eq:v29}
\end{equation}
where $\eta(t)$ is a Gaussian random variable, corresponding to white noise, with correlator
\begin{equation}
\langle \eta(t) \eta(t^{\prime}) \rangle = \Gamma \delta(t-t^{\prime}) \label{eq:v30}
\end{equation}
We do not assume a fluctuation-dissipation relation and take $\gamma$ and $\Gamma$ as independent parameters.

\begin{figure}[htbp]
   \centering
   \includegraphics[width=3in]{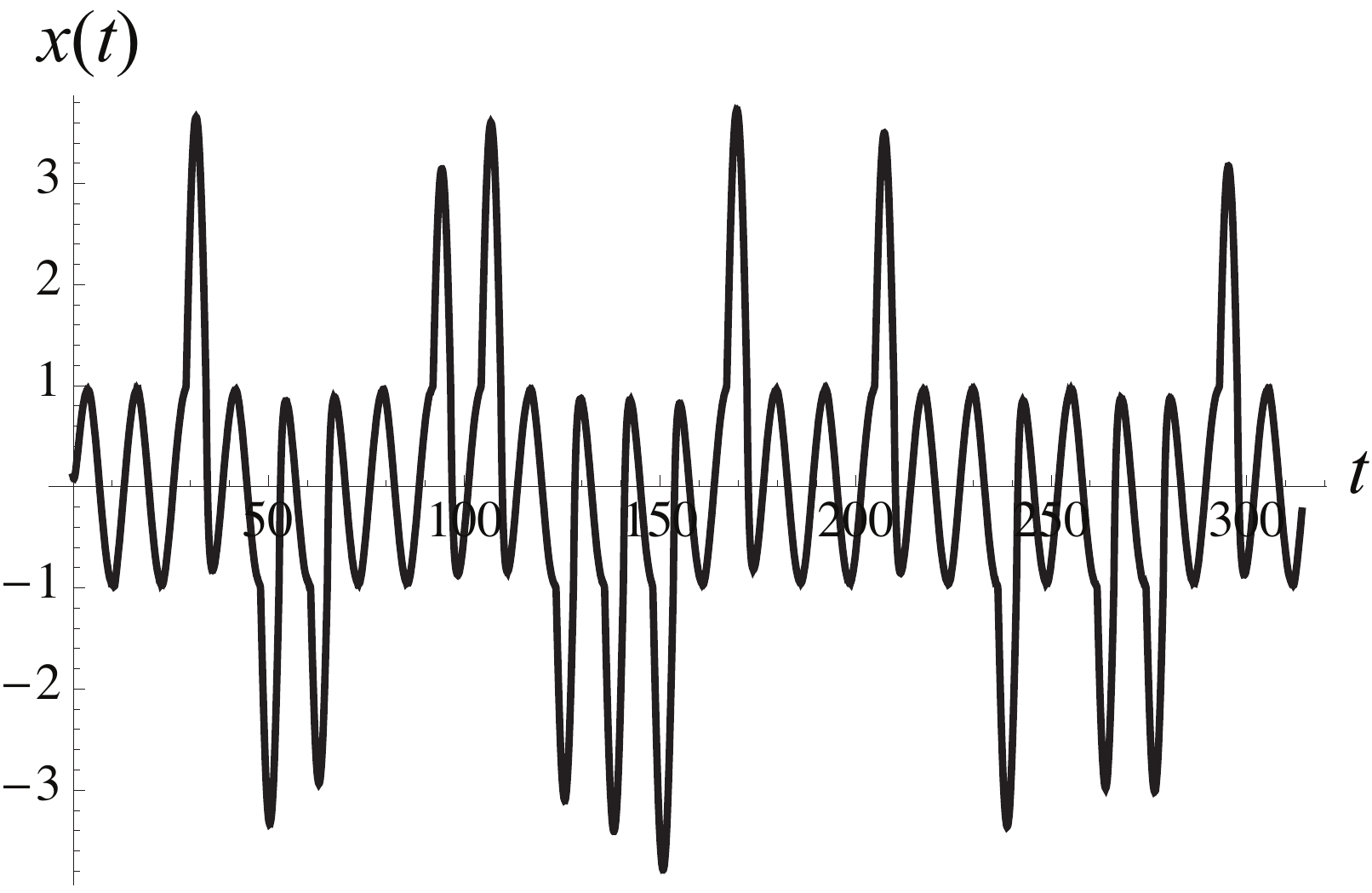}
   \caption{One realization of the solution to the stochastic equation (\ref{eq:v29}) when the noise-free system is in the vicinity of a first-order dynamical phase transition from a phase with a single symmetric solution to a phase with three stable (noise free) steady state solutions, one symmetric about $x=0$ and a pair of skewed solutions. The restoring force is provided in Appendix \ref{app:restore} with $\alpha=0.01, k=2, F_0=0$. Other parameters: drive amplitude, $A=1.89$, frequency, $\omega=0.5$, noise correlator,  $\Gamma=0.0001$.}
   \label{fig:noise1}
\end{figure}

A simple Euler forward integration leads to
\begin{equation}
x(t + \delta t) = x(t) + \frac{\delta t}{\gamma}  [-V^{\prime}(x(t)) + A \sin ( \omega t)] + \sqrt{\Gamma \delta t}  \, y_i \label{eq:v31}
\end{equation}
where $y_i$ is a Gaussian random variable with zero mean and unit standard deviation. As an example, we choose parameters for which the noise-free system has a first order transition from a ``phase" at sufficiently low driving amplitude $A$ in which there is a single, stable steady state, which is symmetric, to a phase (probably not physically reachable, occupying a tiny region of parameter space) at higher driving amplitude with three stable solutions, one symmetric and a pair of skewed solutions and, finally, at even higher amplitude drive to a phase in which only the skewed solutions are stable. For reference, for the parameters we choose, the noise-free dynamical transition occurs with amplitude about $A\simeq1.89$. A realization of the three-solution situation with added noise is shown for $A=1.89$ in Fig. \ref{fig:noise1}.

The noise level $\Gamma = 10^{-4}$ has been chosen so that the noise-driven switching between the stable solutions is apparent. The nature of the switching will be considered in future work. With this level of noise, the single symmetric solution below the transition is unremarkable and not shown.

For modest levels of noise, such as those considered above, qualitatively one expects the sharp noise-free transitions to be rounded, but to retain hints of the underlying noise-free transitions. This is, in fact realized. However, the calculation turns out to be more straightforwardly carried out in the context of the Fokker-Planck equation, to which we turn now.

\subsection{Master equation}\label{sec:FPE}
\begin{figure}[htbp]
\begin{center}
\includegraphics[width=3in]{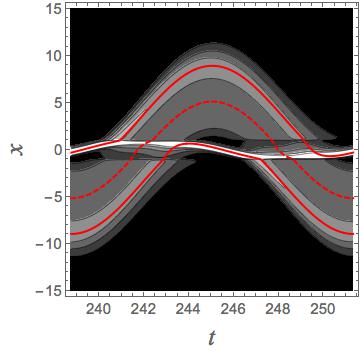}
\caption{ One period (the twentieth) of the probability distribution, $P(x,t)$ as given by the solution of the Fokker-Planck equation (\ref{eq:FPE1})--(\ref{eq:FPE3}), with $A=2.7$, $\omega=0.5$, $k=2$, $F_0=0$ and $\Gamma=0.2$. Also shown are the two stable skewed solutions (solid red curves) and the unstable symmetric solution (dashed red curve). Lighter colors correspond to higher values of $P(x,t)$.}
\label{fig:FPE_steadystate}
\end{center}
\end{figure}

The dynamical equation with noise, Eq. \ref{eq:v29} is Markovian, of the form $\dot{x}=h(x,t)+\eta(t)$, where, as above, $\eta(t)$ is delta-correlated with {\it constant} $\Gamma$, and $h(x,t)$ contains the ``systematic" restoring force and the time-dependent driving force. As shown, e.g., in \cite{Risken}, one can reformulate the analysis of this equation into a probability distribution, $P(x,t)$, of an ensemble of solutions to the noisy equation of motion. This distribution is governed by the Fokker-Planck equation,
\begin{eqnarray}
\frac{\partial}{\partial t} P(x,t) &=& -\frac{\partial}{\partial x}D^{(1)}(x,t) P(x,t)  \nonumber \\
 {} & & + \left(\frac{\partial}{\partial x}\right)^2 D^{(2)}(x,t) P(x,t) \label{eq:FPE1}    \\
D^{(1)}(x,t) &=& h(x,t) \label{eq:FPE2} \\
D^{(2)}(x,t) &=& \Gamma/2 \label{eq:FPE3}
\label{eq:fpe}
\end{eqnarray}
In our case, the quantity $h(x,t)$ is the driving force $F(x(t)) + A \sin(\omega t)$. As an example of the utilization of this equation, we apply it to the case of a discontinuous transition described in Sec. \ref{sec:transnature}. Initially, we take $\Gamma = 0.2$, with the amplitude, $A$ equal to 2.7.
\begin{figure}[htbp]
\begin{center}
\includegraphics[width=3in]{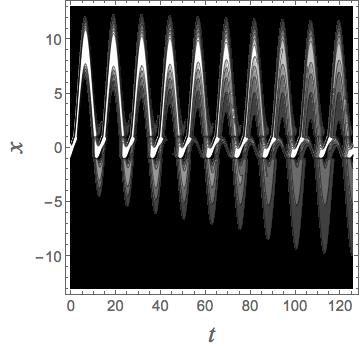}
\caption{The probability distribution, $P(x,t)$ as given by the solution of the Fokker-Planck equation (\ref{eq:FPE1})--(\ref{eq:FPE3}), with $A=2.7$, $\omega=0.5$, $k=2$, $F_0=0$ and  $\Gamma=0.2$. Ten periods of the driving force are shown. Lighter colors correspond to higher values of $P(x,t)$. The distribution at $t=0$ was chosen to center on one of the dynamically stable skewed solutions to the deterministic equation of motion. }
\label{fig:FPE_evolution}
\end{center}
\end{figure}
At this amplitude there are two stable skewed solutions and one unstable symmetric solution. The initial distribution was chosen to center on one of the skewed solutions. As illustrated in Fig. \ref{fig:FPE_evolution} the distribution evolves toward one that is symmetric. That is, both skewed solutions are equally represented in the steady state distribution. In fact, one can readily demonstrate that if $P(x,t)$ is a solution to Eqs. (\ref{eq:FPE1})--(\ref{eq:FPE3}), then so is $P(-x,t+ \pi/\omega)$. In Fig. \ref{fig:FPE_steadystate}, we show the steady state distribution for one period, along with the two stable skewed solutions and the unstable symmetric solution for zero noise.

\subsection{Compliances in the presence of noise} \label{sec:noisecomplainces}
\begin{figure}[htbp]
\begin{center}
\includegraphics[width=3in]{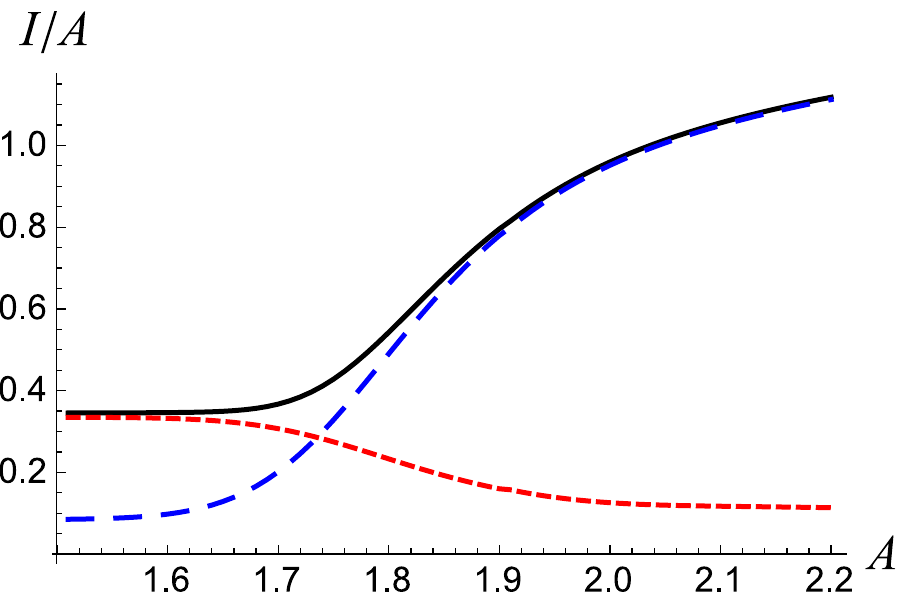}
\caption{The total compliance $I_t/A$ (solid black curve), the dissipative compliance, $I_{d}/A$ (long-dashed blue curve) and the storage compliance, $I_s/A$ (dashed red curve). Here $k=2$, $\omega=0.5$, $F_0=0$ and $\Gamma=0.014$. }
\label{fig:noisecomp}
\end{center}
\end{figure}

Making use of the steady state probability distribution, one can calculate the effect of noise on compliances. Figure \ref{fig:noisecomp}, displays the results for the compliances in the case in which the parameters in the noise-free equation of motion are such that Eq. (\ref{eq:eom1}) predicts a first order transition from a dynamically stable symmetric solution to stable skewed solutions. In this calculation, $\Gamma$ is taken to be $0.014$---significantly higher values of this quantity lead to a washing out of evidence of a dynamical transition. Here the compliances are defined as in Sec. \ref{sec:compliances}, with $x(t)$ replaced by $\langle x(t) \rangle$, the average being taken with respect to the distribution function $P(x,t)$, i.e. $\langle x(t) \rangle = \int_{-\infty}^{\infty} x^{\prime}P(x^{\prime},t) dx^{\prime}$.
The plot in Fig. \ref{fig:noisecomp} is to be compared with Fig. \ref{fig:phaset2}, the plot of compliances in the absence of noise. In the presence of noise the transition from an elastic to a viscous response regime no longer displays a first order discontinuity.

\section{Concluding remarks} \label{sec:concluding}
The investigations of Zocchi \textit{et al. } \cite{zocchi1,zocchi2,zocchi4,zocchi3} of the response of a folded protein to a sinusoidal driving force point to a description in terms of a single collective coordinate. Moreover, the response appears to be quantifiable in terms of a viscoelastic formalism, with the additional feature of a sharp transition from dominantly elastic response at low amplitude drive to mainly viscous response at high drive amplitude. We find that a simple model exhibits behavior consistent with those observations. From such analysis we  conclude that the combination of experimental findings and characterization in terms of a simple model by Zocchi et al. forms a promising basis for a comprehensive description of the gross mechanical and dynamical properties of a class of folded proteins. Furthermore, we strongly believe that it ought to be useful in further investigations, both experimental and theoretical, of those properties.

Extensions of the present study should aid in the development of insights into the structural properties of proteins, both in isolation and as components of larger systems. For example, it would be interesting to expand the characterization of a protein configuration beyond a single collective variable, to take into account internal adjustments to external influences. Given the size and complexity of those molecules the number of mechanical degrees of freedom will be considerable. However, judicious analysis should allow the investigator to expand the set of dynamical variables to a manageable size. Additionally, the effects of inertia, while small, are almost certainly non-negligible, and the effects of noise deserve fuller attention.

The existence of a dynamical phase transition  in the simple model studied here deserves note in and of itself, and the genesis of this transition merits further study. It would be of great interest to identify the essential characteristics of a nonlinear system that lead to this phenomenon. As regards symmetry breaking, a crucial aspect of the model is the existence of a symmetry in the underlying equation of motion, in particular the fact that the restoring force, $F(x)$ satisfies $F(-x)=-F(x)$; a similar symmetry is present in the Suzuki-Kubo model \cite{sk,TandO}. However,  the protein conformation is almost certainly not consistent with such a force, given that the collective co-ordinate $x$ will have one sign---say positive---when the protein is stretched and the opposite sign when it is compressed. It is entirely reasonable to expect that the ``cracking'' process, in which a linear restoring force is replaced by another relationship between that force and the relative displacements in the protein, will differ in those two regimes, which means that there is no symmetry to break. Nevertheless, as shown in Appendix \ref{app:skewed}, even in the absence of such intrinsic asymmetry, the equation of motion can nevertheless lead to the kind of dynamical transition discussed in this paper.  Thus, there are good reasons to expect that the response observed by Zocchi and co-workers does indeed arise from a true dynamical phase transition.

\begin{acknowledgements}

We are grateful to Profs. Giovani Zocchi and Jonathan Rubin for interesting and helpful discussions. JR acknowledges support from the National Science Foundation through DMR Grants 1006128 and 1309423.

\end{acknowledgements}

\begin{appendix}

\section{The nonlinear restoring force} \label{app:restore}

The form for the restoring force derives from the function
\begin{equation}
\Delta(x, \alpha) =\frac{1}{\pi} \frac{\alpha}{x^2 + \alpha^2} \label{eq:restore1}
\end{equation}
In the limit $\alpha =0$, this is just the Dirac delta function. Two functions that result from integrating this function once and then twice are
\begin{eqnarray}
f_1(x, \alpha) & = & \frac{2 \tan ^{-1}\left(x/\alpha\right)}{\pi } \label{eq:restore2} \\
f_2(x, \alpha) & = &\frac{2 \left(x \tan ^{-1}\left(x/\alpha \right)-\frac{1}{2} \alpha
   \ln \left(\alpha ^2+x^2\right)\right)}{\pi } \nonumber \\  \label{eq:restore3}
\end{eqnarray}

\begin{figure}[htbp]
\begin{center}
\includegraphics[width=3in]{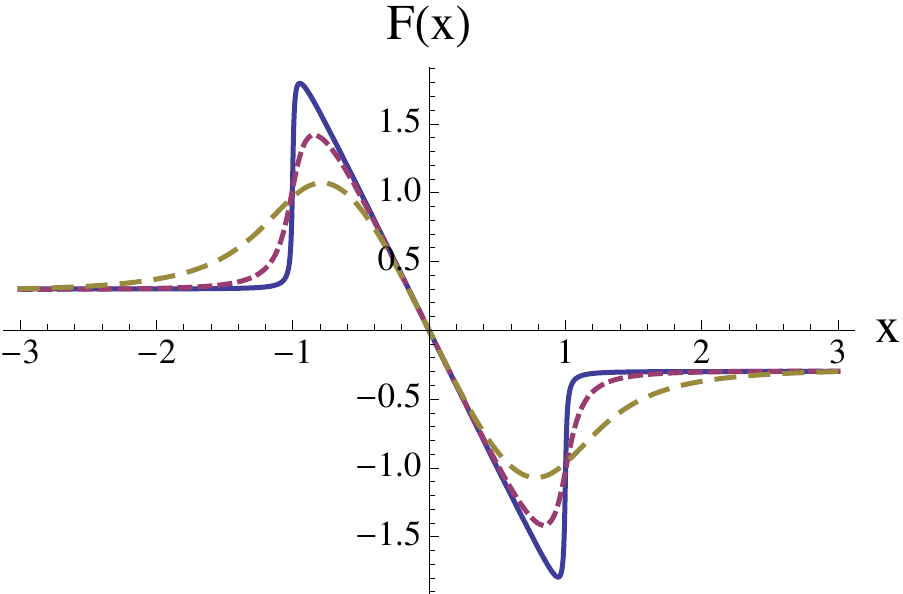}
\caption{Three instances of restoring forces arising from (\ref{eq:restore6}). In all cases, the absolute value of the slope of the curve at $x=0$ is $k=2$, and the absolute value of the asymptote is $F_0=0.3$. The parameter $\alpha$ controlling the transition between the central and outer region is 0.01 for the solid curve, 0.1 for the short-dashed curve and 0.5 for the long-dashed curve. The restoring force changes from Hookean to constant near $x=\pm1$. }
\label{fig:3Forces}
\end{center}
\end{figure}

Then, with coefficients
\begin{eqnarray}
A & = & -\frac{2 \alpha  F_0 - \pi k(1+  \alpha ^2) }{4 \left((1+\alpha ^2) \tan
   ^{-1}\left(1/\alpha \right)-\alpha \right)} \nonumber \\ && -\frac{F_0}{2} \label{eq:restore4} \\
B & = & -\frac{2 \alpha  F_0 -\pi k(1+ \alpha ^2) }{4 \left((1+\alpha ^2) \tan
   ^{-1}\left(1/\alpha \right)-\alpha \right)} \label{eq:restore5}
\end{eqnarray}
The function
\begin{eqnarray}
\lefteqn{A \left(f_1(x+1, \alpha)+f_1(x-1,\alpha) \right)} \nonumber \\ && + B \left( f_2(x-1,\alpha)-f_2(x+1,\alpha) \right) \label{eq:restore6}
\end{eqnarray}
has the form shown in Fig. \ref{fig:enforce}, with slope $-k$ at the origin and asymptotes of $\pm F_0$. The parameter $\alpha$ determines the sharpness of the transition from a linear restoring force to a restoring force that is constant. This cumbersome expression for the restoring force enables independent variation of the features controlled by $\alpha$ and $F_0$. Figure \ref{fig:3Forces} displays three instances of the kind of restoring force that we explore.

In  the body of the paper we have taken the transition from Hookean behavior  to occur around $|x| =1$.

\section{Pairs of skewed solutions} \label{app:pairs}

The form of the equation of motion, (\ref{eq:eom1}) and the symmetry of the restoring force,
\begin{equation}
F(x) =-F(-x) \label{eq:pair1}
\end{equation}
allow one to generate, from a solution $x(t)$, a ``mirror image'' solution equal to $-x(t+ \pi/\omega)$. To see this, replace $t$ by $t+ \pi/\omega$ in (\ref{eq:eom1}), which leads to
\begin{eqnarray}
\frac{dx(t + \pi/\omega)}{dt} &=& \frac{1}{\gamma} \left[F(x(t+ \pi/\omega)) + A \sin(\omega(t+ \pi/\omega)) \right] \nonumber \\
& = & \frac{1}{\gamma} \left[ -F (-x(t  + \pi/\omega)) - A \sin (\omega t) \right] \label{eq:pair2}
\end{eqnarray}
If we now multiply both sides of the resulting equation by $-1$, we end up with the result that $-x(t + \pi/\omega)$ satisfies the same equation as $x(t)$. This allows one to generate, from any skewed solution, a second one, skewed in the opposite direction. By contrast, a symmetric solution will reproduce itself under this transformation.

\section{Stability analysis for solutions to the equation of motion in Sec. \ref{sec:analytical} }\label{app:stability}
To assess the stability of a solution of the equation of motion when the restoring force exhibits a sharp break between Hooke's Law and vanishing amplitude as in (\ref{eq:forcelim1}), we perturb that equation as follows. Assume a solution, $x_0(t)$, to (\ref{eq:eom1}). Then, write $x(t) =x_0(t) + \delta x(t)$. At first order in the perturbation $\delta x(t)$, the equation becomes
\begin{equation}
\frac{d \delta x(t)}{dt} = \frac{1}{\gamma} \left[\delta x(t) F^{\prime}(x_0(t)) \right] \label{eq:sa1}
\end{equation}
Dividing both sides of (\ref{eq:sa1}) by $\delta x(t)$ and integrating over a period of the drive, we find
\begin{equation}
\ln \left(\frac{\delta x (t+ 2 \pi/\omega)}{\delta x(t)} \right) = \frac{1}{\gamma} \int_t^{t+2 \pi/\omega} F^{\prime}(x_0(t)) dt \label{eq:sa2}
\end{equation}
or
\begin{equation}
\frac{\delta x (t+ 2 \pi/\omega)}{\delta x(t)} = \exp \left[ \frac{1}{\gamma}\int_t^{t+ 2 \pi/\omega}F^{\prime}(x_0(t) ) dt \right] \label{eq:sa3}
\end{equation}

\begin{figure}[htbp]
\begin{center}
\includegraphics[width=3in]{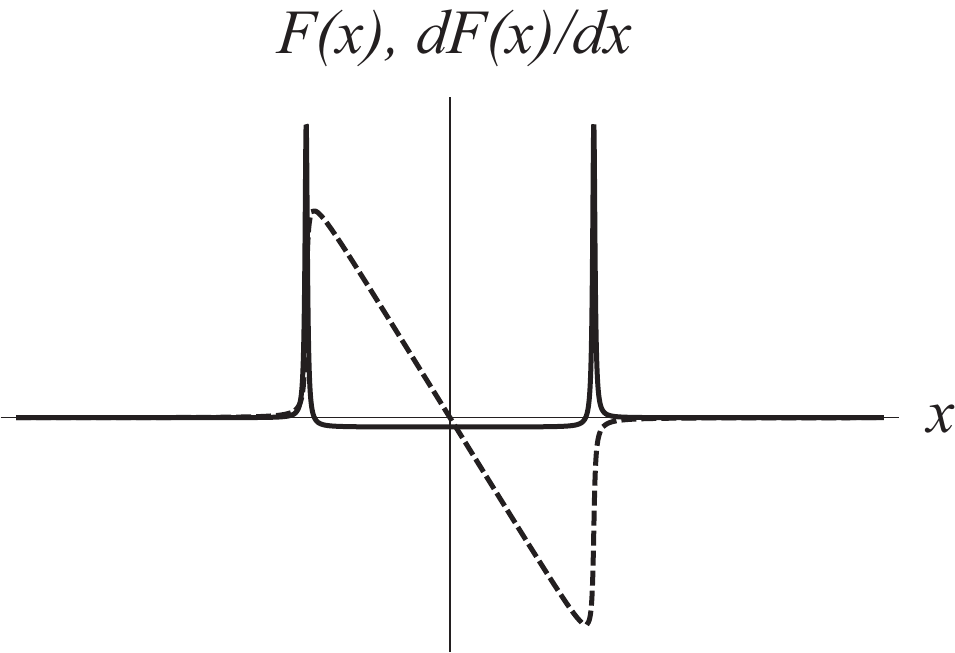}
\caption{Restoring force with a rapid transition between Hooke's law in a region about the origin (dashed curve) along with its derivative (solid curve). When the transition is perfectly abrupt the peaks in the force derivative become Dirac delta functions. }
\label{fig:forceandderivative}
\end{center}
\end{figure}

Figure \ref{fig:forceandderivative} shows a restoring force that changes abruptly from Hookean to vanishing, along with its derivative.
In the limit of a discontinuous transition, the peaks in the derivative become Dirac delta functions. If we are interested in the skewed solutions described in Sec. \ref{sec:analytical}, then the integral on the right hand side of (\ref{eq:sa2}) and (\ref{eq:sa3}) yields three contributions. The first is from the interior region and will be equal to $-k$ multiplied by the amount of time that $x(t) < x_0$, which, according to (\ref{eq:window}), is $2 \pi/\omega -2t_0$. Then there is the result of integrating through the very narrow peaks in the derivative. Here, we write
\begin{eqnarray}
\int F^{\prime}(x_0(t)) dt &= & \int \frac{F^{\prime}(x_0(t))}{dx_0(t)/dt} dx_0(t) \nonumber \\
& = & \int \frac{F^{\prime}(x)}{ F(x) \pm A \sin(\omega t_0)}dx \nonumber \\
& \rightarrow & \pm \ln \left[ \frac{A \sin (\omega t_0)}{A \sin (\omega t_0) \mp kx_0}\right] \label{eq:sa4}
\end{eqnarray}
The upper sign on the right hand side of (\ref{eq:sa4}) applies in the case that the trajectory through the peak in the derivative is out of the inner region and the lower sign applies when the trajectory is into the inner region. The time, $t$ has been set equal to $t_0$, as passage through the peaks occurs in effectively a single instant of time.  Assembling the two contributions we fnd for the right hand side of (\ref{eq:sa3})
\begin{eqnarray}
\exp \left[-\frac{k}{\gamma}\left(\frac{2 \pi}{\omega}-2t_0 \right) + \frac{1}{\gamma} \ln \left(\frac{ (A \sin (\omega t_0))^2}{(A \sin( \omega t_0))^2-(kx_0)^2}\right)\right] \nonumber \\  \label{eq:sa5}
\end{eqnarray}
If the expression in square brackets is real and negative then perturbations to the trajectory $x_0(t)$ die off, and the periodic solution to the equation of motion is stable. If it is positive, then those perturbations grow exponentially, and the solution is unstable. If the expression is imaginary, then the assumption of a solution is incorrect, in that the velocity at one of the transitions between Hooke's law force and no restoring force has the wrong sign. We find that in the case of graphical solutions to Eq. (\ref{eq:v9}) that are not equal to $\pi/\omega$, the one with a smaller value of $t_0$ does not correspond to an actual trajectory, and the one with the larger value of $t_0$ corresponds to a dynamically stable solution to the equation of motion.

\section{Review of the statistical mechanics of tricriticality} \label{app:tricrit}

\begin{figure}[htbp]
\begin{center}
\includegraphics[width=3in]{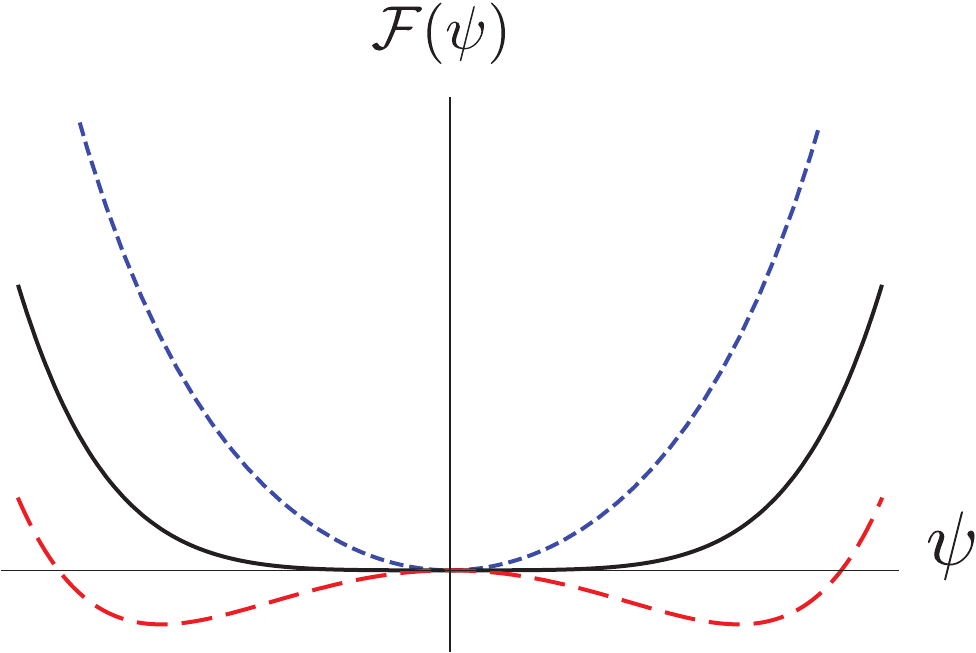}
\caption{Behavior of the free energy $\mathcal{F}(\psi)$ in (\ref{eq:tc1}) when the coefficients $u$ and $v$ are positive. The blue, dashed curve is the free energy when $t>0$. The red, long-dashed, curve is the free energy when $t<0$. The solid black curve is $\mathcal{F}(\psi)$ at the transition point, when $t=0$. }
\label{fig:tc1}
\end{center}
\end{figure}

According to the canonical model of tricritical points based on an order parameter, the thermodynamic behavior of a system near a symmetry breaking phase transition is governed by a free energy of the form
\begin{equation}
\mathcal{F(\psi)} = \mathcal{F}_0 + Ct \psi^2 + u \psi^4 + v \psi^6 \label{eq:tc1}
\end{equation}
Here, $\psi$ is the order parameter and $t$ is a reduced temperature, proportional to the difference between the actual temperature of the system and a critical temperature. The coefficients $C>0$, $u$ and $v$ are assumed to be insensitive to the temperature as long as $t$ is sufficiently small. Thermodynamic stability requires $v > 0$; otherwise, the free energy takes on unbounded negative values as $|\psi|$ increases. The system described by this free energy settles into the state of lowest free energy, determined by minimizing $\mathcal{F}( \psi)$. The results of this minimizing procedure depend qualitatively on the sign of the coefficient $u$. When $u$ is positive, the $\psi$-dependence of $\mathcal{F}$ is as shown in Fig. \ref{fig:tc1}.
As shown in the figure, when $t>0$, the unique minimum of the free energy is at $\psi=0$. On the other hand, when $t<0$, there are two minima, spaced equidistant from $\psi=0$. The minimum at $\psi=0$ has become a free energy maximum. In mechanics, that maximum would correspond to a point of unstable equilibrium, in contrast to the minima, which represent points of stable equilibrium. However, a thermodynamic system will inevitably fluctuate away from such a state. Because the free energy is even in $\psi$, in that $\mathcal{F}(- \psi) = \mathcal{F}(\psi)$, each of the minima for $t<0$, which lies to one side or the other of the origin, represents a violation of the symmetry of the physics underlying the system's thermodynamics, and the transition that occurs as $t$ passes through zero is called a \textit{symmetry breaking} phase transition. The determination of the minima follows from the solution of the equation
\begin{equation}
\frac{d \mathcal{F}(\psi)}{d \psi} =0 \label{eq:tc2}
\end{equation}

\begin{figure}[htbp]
\begin{center}
\includegraphics[width=3in]{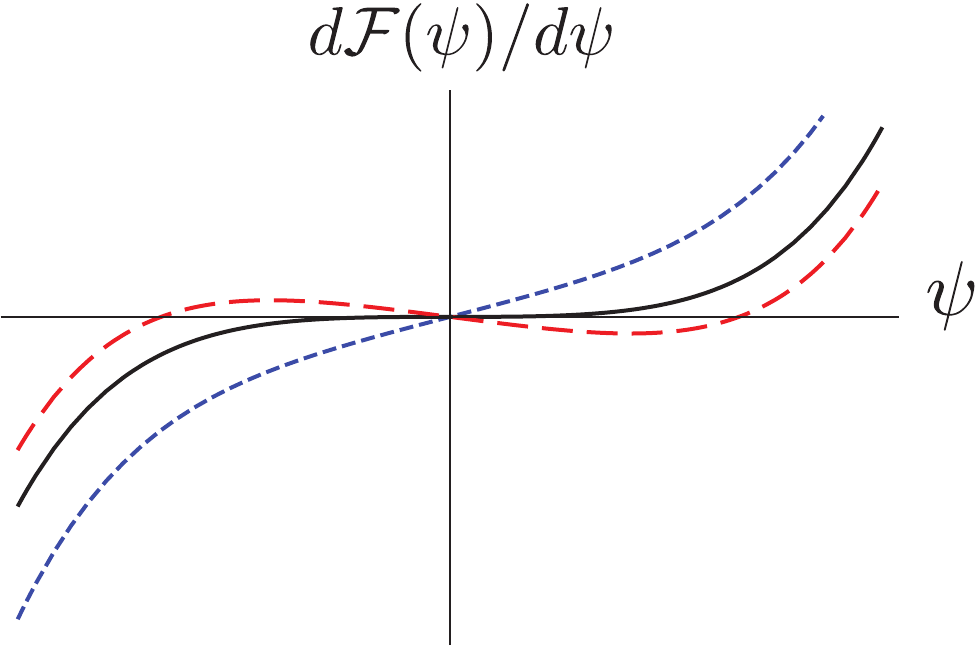}
\caption{The graphical representation of $d \mathcal{F}( \psi)/d \psi$. The curves follow the format of Fig. \ref{fig:tc1}. }
\label{fig:tc1a}
\end{center}
\end{figure}

\begin{figure}[htbp]
\begin{center}
\includegraphics[width=3in]{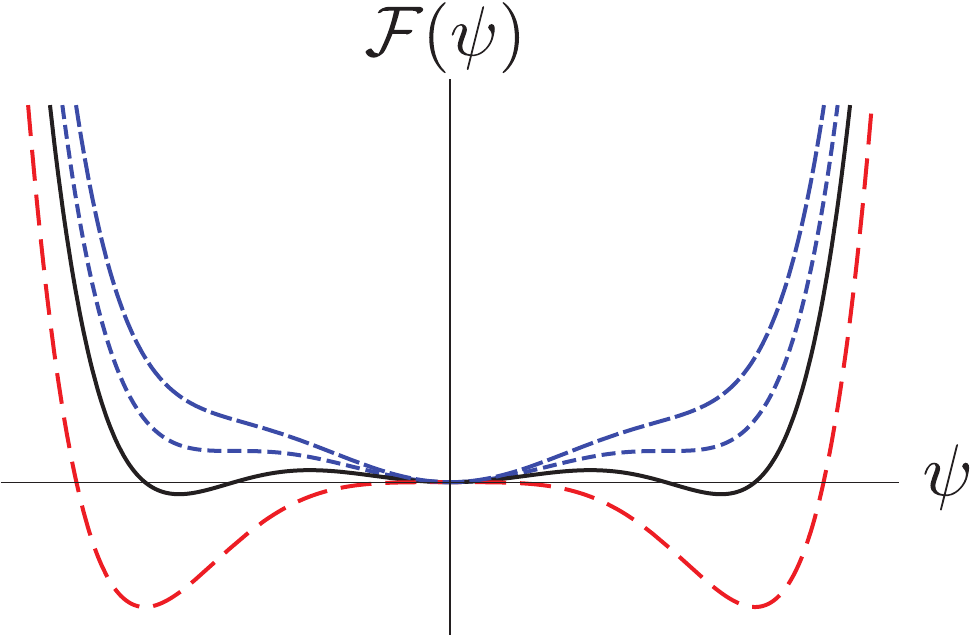}
\caption{Curves for $\mathcal{F}( \psi)$ when the fourth order coefficient $u$ is negative. At high enough temperatures (blue long dashed curve) there is one free energy minimum, at $\psi=0$. As the temperature is lowered, two other minima emerge (blue dashed curve and solid black curve). In addition, two maxima that separate the minimum at $\psi=0$ and the flanking minima appear. At low enough temperatures the two minima at non-zero $\psi$ remain, the central extremum in the free energy having become a maximum (red long-dashed curve).}
\label{fig:tc2}
\end{center}
\end{figure}

\begin{figure}[htbp]
\begin{center}
\includegraphics[width=3in]{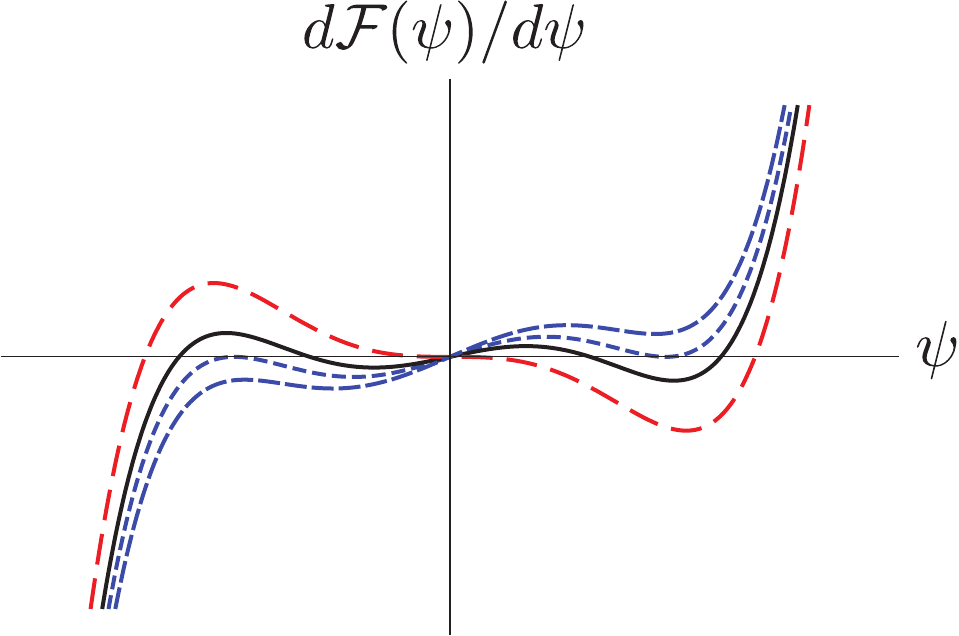}
\caption{ Curves for $d \mathcal{F}( \psi)/d \psi$ corresponding to the curves shown in Fig. \ref{fig:tc1}. These curves pass through zero at extrema---maxima and minima---of $\mathcal{F}( \psi)$. They are formatted in the same way as the free energy curves in Fig. \ref{fig:tc2}.}
\label{fig:tc2a}
\end{center}
\end{figure}

On the other hand, when $u <0$, the behavior of $\mathcal{F}(\psi)$ is shown in Fig. \ref{fig:tc1a}.
In this case, the minima at non-zero values of $\psi$ develop while there is still a free energy minimum at $\psi=0$. As that occurs two free energy maxima appear between the new minima and the free energy minimum at $\psi=0$. Eventually, at low enough temperatures, the only minima are at finite $\psi$, with one free energy maximum at $\psi=0$. Free energies and the related curves for the free energy derivative are displayed in Figs. \ref{fig:tc2} and \ref{fig:tc2a}.
Focusing on the curves in Fig. \ref{fig:tc2a}, we see that at high enough temperatures, there is only one value of $\psi$ at which $d \mathcal{F} (\psi)/ d \psi$ passes through zero: $\psi=0$. As the temperature is lowered additional zeros appear, first four, two on each side of the origin, corresponding to a maximum and then a minimum in $\mathcal{F}( \psi)$. At even lower temperatures, only three zeros remain, corresponding to minima at non-zero values of $\psi$, while the zero at $\psi=0$ refers to a free energy maximum at that point.

\section{Alternate stability analysis of steady state solutions to (\ref{eq:in1})} \label{app:instability}

As an alternate approach to the stability analysis of steady state solutions to (\ref{eq:in1}), obtained by determining simultaneous solutions to (\ref{eq:in6}) and (\ref{eq:in7}), we return to the original equation (\ref{eq:in1}) and expand about periodic solution, $x_0(t)$. That is, write $x(t) = x_0(t) + \delta x(t)$. The equation satisfied by $\delta x(t) $ is
\begin{equation}
\frac{d^2 \delta x(t)}{dt^2} + \gamma \frac{d\delta x(t)}{dt} = F^{\prime}(x_0(t)) \delta x(t) \label{eq:in12}
\end{equation}
Then, we write $\delta x(t) = \psi(t) e^{- \gamma t/2}$. The equation for $\psi(t)$ is
\begin{equation}
\frac{d^2 \psi(t)}{dt^2} - \left(\frac{\gamma}{2} \right)^2 \psi(t) = F^{\prime}(x_0(t)) \psi(t) \label{eq:in13}
\end{equation}
Because of the periodicity of $x_0(t)$, this is just like the Schr\"{o}dinger equation of a particle in a periodic potential. The periodicity is that of the driving force. We can now treat this as a problem in one dimensional band theory \cite{AandM}. We start with a solution of the above equation, $\psi_1(t)$ that has the initial conditions
\begin{eqnarray}
\psi_1(0) & = & 1 \label{eq:in14} \\
\psi_1^{\prime}(0) & = & 0 \label{eq:in15}
\end{eqnarray}
and a function $\psi_2(t)$ satisfying
\begin{eqnarray}
\psi_2(0) & = & 0 \label{eq:in16} \\
\psi_2^{\prime}(0) & = & 1 \label{eq:in17}
\end{eqnarray}
A general solution will be a linear combination of $\psi_1(t)$ and $\psi_2(t)$, where the coefficients are determined, for example, by the value of the solution and the value of its derivative at $t=0$. If we write
\begin{equation}
\psi(t) = A \psi_1(t) + B \psi_2(t) \label{eq:in18}
\end{equation}
Then,
\begin{eqnarray}
A & = & \psi(0) \label{eq:in19} \\
B & = & \psi^{\prime}(0) \label{eq:in20}
\end{eqnarray}
After one full period, the magnitude and derivative of the solution is given by
\begin{eqnarray}
A^{\prime} & = & A \psi_1(2 \pi/\omega) + B \psi_2(2 \pi/\omega) \label{eq:in21} \\
B^{\prime} & = & A \psi_1^{\prime}(2 \pi/\omega) + B \psi_1^{\prime}(2 \pi/\omega) \label{eq:in22}
\end{eqnarray}
This gives rise to the map
\begin{equation}
\left( \begin{array}{l} A^{\prime} \\ B^{\prime} \end{array}\right) = \left(\begin{array}{ll} \psi_1(2 \pi/\omega) & \psi_1 (2 \pi/\omega) \\ \psi_1^{\prime}(2 \pi/\omega) & \psi_2^{\prime}(2 \pi/\omega)\end{array} \right) \left( \begin{array}{l} A \\ B \end{array}\right) \label{eq:in23}
\end{equation}
We know from the invariance of the Wronskian and from its value at $t=0$ that
\begin{equation}
\psi_1(t) \psi_1^{\prime}(t) - \psi_1^{\prime}(t) \psi_2(t) =1 \label{eq:in24}
\end{equation}
for all $t$. This means that the determinant of the matrix in (\ref{eq:in23}) is equal to one.

Given that the relationship (\ref{eq:in23}) holds for the values of $A$ and $B$ for each succeeding period, the ultimate behavior of the solution is going to be determined by the eigenvalues of the matrix on the right hand side of that equation. In light of the relationship between $\delta x(t)$ and $\psi(t)$, we see that the stability of the solution is going to be determined by the exponential $e^{- \gamma \pi/\omega}$ and the eigenvalues. Specifically, by the product
\begin{eqnarray}
\lefteqn{e^{- \gamma \pi/\omega} \frac{1}{2} \Bigg[\psi_1( 2 \pi/\omega) + \psi_2^{\prime}(2 \pi/\omega) }\nonumber \\ &&  \pm \sqrt{(\psi_1( 2 \pi/\omega) + \psi_2^{\prime}(2 \pi/\omega))^2-4}\Bigg] \label{eq:in25}
\end{eqnarray}
If the absolute value of (\ref{eq:in25}) is less than one, for both signs in the expression, then the solution is stable, and if that is not the case the solution is unstable. The result in (\ref{eq:in25}) follows in part from the fact that the determinant of the matrix in (\ref{eq:in23}) is equal to one.

Making use of the above method, we arrive at the same conclusions regarding the steady state solutions to the equation of motion (\ref{eq:in1}) shown in Sec. \ref{sec:inertia}.

\section{Dynamical transition when the force is inherently asymmetric} \label{app:skewed}

\begin{figure}[htbp]
\begin{center}
\includegraphics[width=3in]{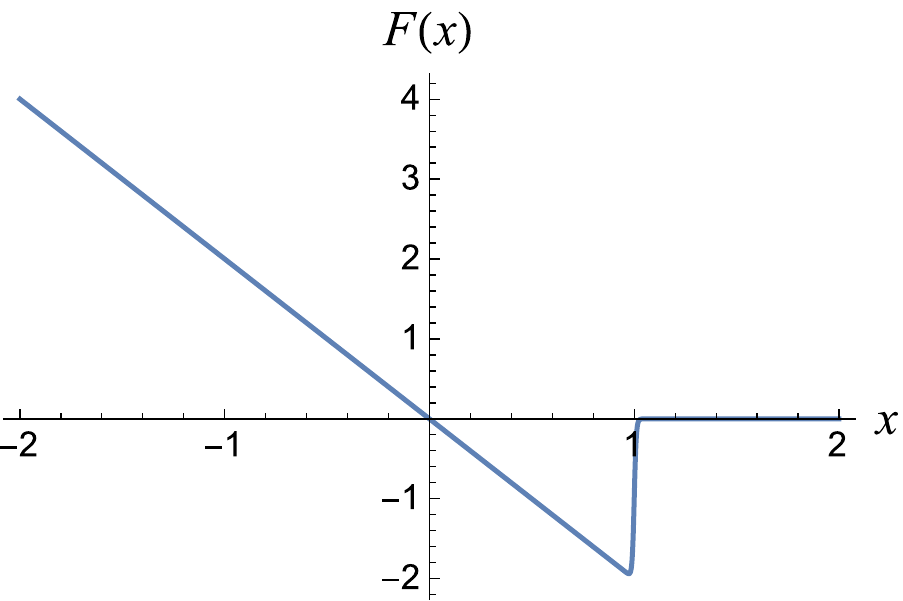}
\caption{A restoring force, $F(x)$ that is Hookean for all positive values of $x$ but that reverts to zero at $x=-1$. The spring constant in the Hookean regime is $k=2$. }
\label{fig:sk1}
\end{center}
\end{figure}

We have been considering a system in which the force, $F(x)$ that restores it to stable equilibrium in the absence of an external drive is symmetric in the dynamical variable that characterizes its distortion from that state in that $F(-x) = -F(x)$.  It is natural to ask whether this symmetry in the equation of motion is essential to the existence of the dynamical transitions discussed here. As it turns out those transitions can also arise when the underlying symmetry is absent. Consider the restoring force shown in Fig. \ref{fig:sk1}.
As shown in that figure, the force is of the form $F=-kx$, with spring constant $k=2$ for $x <1$. Outside of that region the restoring force is equal to zero.

\begin{figure}[htbp]
\begin{center}
\includegraphics[width=3in]{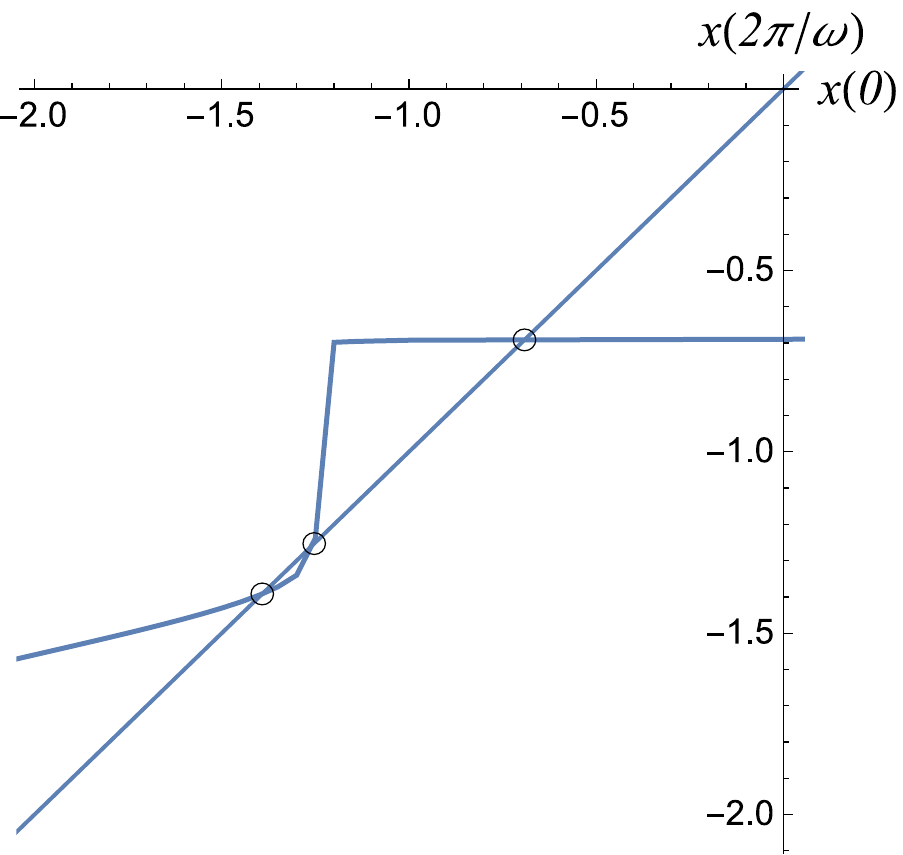}
\caption{The map from $x(0)$ to $x(2 \pi/ \omega)$ generated by the equation of motion (\ref{eq:eom1}) with the restoring force shown in Fig. \ref{fig:sk1} and $\omega=2$, $\gamma=1$ and $A=2.762$. The points of intersection between the map and the $45^{\circ}$ line, also shown, are indicated by  small circles. The two outer intersections correspond to dynamically stable steady state solutions to the equation of motion, while the inner intersection corresponds to a dynamically unstable steady state solution.}
\label{fig:sk2}
\end{center}
\end{figure}

\begin{figure}[htbp]
\begin{center}
\includegraphics[width=3in]{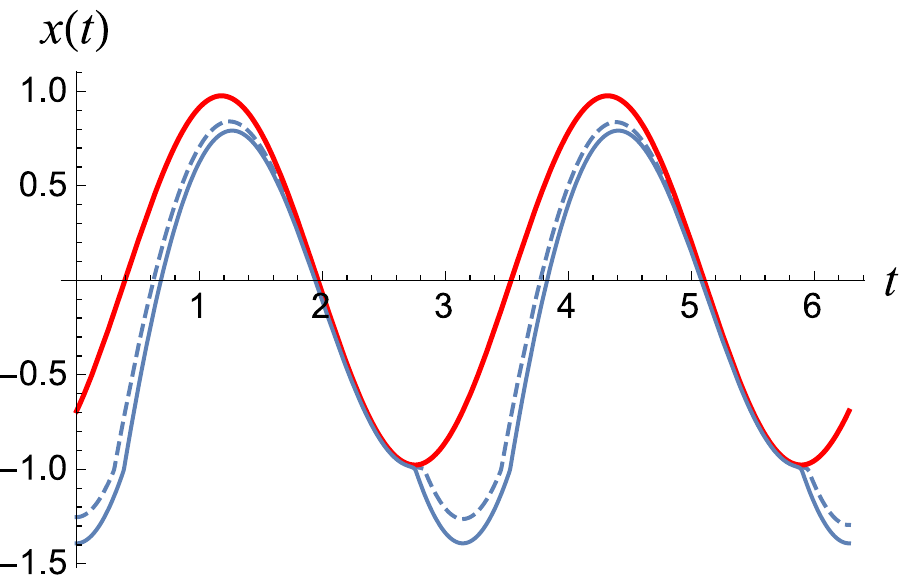}
\caption{The three steady state solutions to the equation of motion corresponding to the points of intersection in Fig. \ref{fig:sk2}, plotted over two periods of the drive, $A \sin(\omega t)$. The thick red curve corresponds to a trajectory that lies entirely in the region in which the restoring force is linear. The thin blue curve is the stable solution that extends outside that region ($x<-1$), and the dashed curve is the dynamically unstable solution that separates the two stable solutions for $x(t)$. }
\label{fig:sk3}
\end{center}
\end{figure}

In Fig. \ref{fig:sk2} we see the map from $x(0)$ to $x( 2 \pi/\omega)$, along with the $45^{\circ}$ line for drive amplitude $A=2.762$, frequency $\omega=2$ and $\gamma=1$ in the equation of motion (\ref{eq:eom1}).
This kind of map is consistent with the existence of two stable fixed point solutions and one unstable solution; Fig, \ref{fig:sk3} shows those three solutions. This coexistence points to a dynamical transition identical to sharp transitions described in the body of this paper, the only difference being the absence of symmetry breaking; there is no underlying symmetry to break.

\begin{figure}[htbp]
\begin{center}
\includegraphics[width=3in]{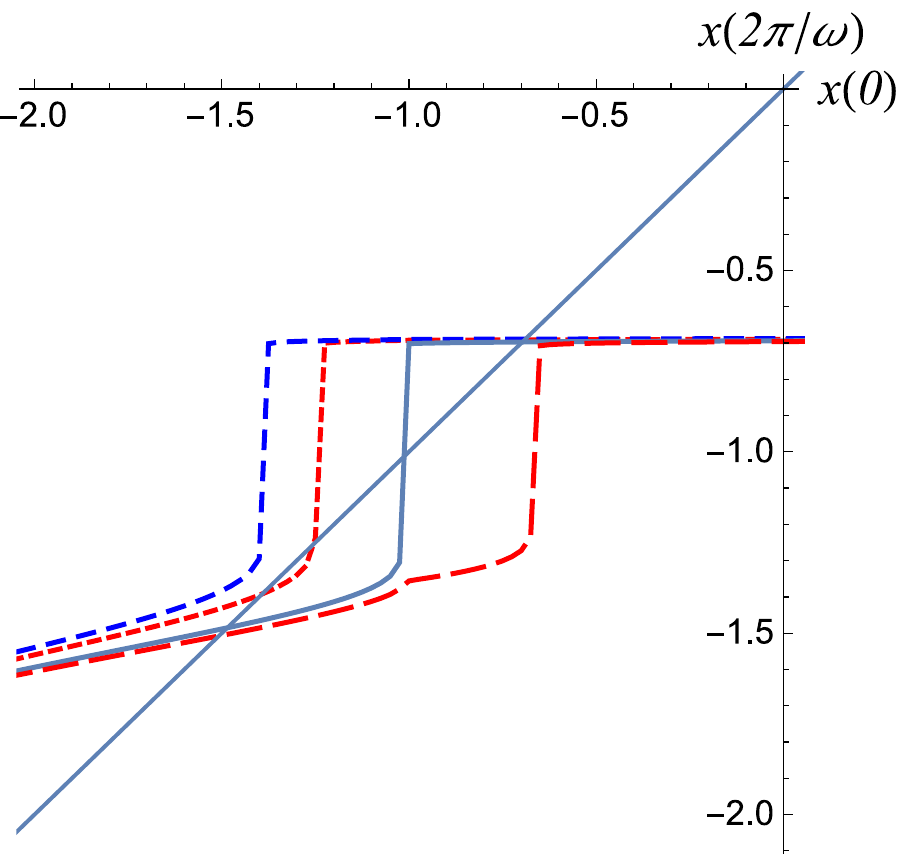}
\caption{Illustrating a first order transition in the case in which the restoring force is as shown in Fig. \ref{fig:sk1}.  The parameters are as described in the caption to Fig. \ref{fig:sk2}, except for the values of the drive amplitudes. Those amplitudes are 2.755 (intermediate dashed blue curve),  2.762 (short dashed red curve), 2.775  (solid black curve) and 2.78 (long dashed red curve). Note that there are no more than three steady state solutions to the equation of motion and that at the extremes there is only one steady state solution to the equation of motion (\ref{eq:eom1}), in contrast to the case of the symmetry breaking dynamical phase transition. }
\label{fig:sk4}
\end{center}
\end{figure}

As further evidence for the dynamical transition, Fig. \ref{fig:sk4} contains the maps for four different drive amplitudes in the same region plotted in Fig. \ref{fig:sk3}. This plot is to be compared with the maps shown in Fig. \ref{fig:fograph}.
The parameters are as described in Fig. \ref{fig:sk2}, except for the drive amplitudes, which range from $A=2.755$ to $A=2.78$. The discussion of the scenario corresponding to this figure parallels the corresponding commentary just above Fig. \ref{fig:fograph}.

The noteworthy results here are, first,  that at the highest drive amplitude there is only one skewed steady state solution to the equation of motion instead of two and, second,  that when there is more than one solution, the two dynamically stable ones correspond to a symmetric solution and a single skewed solution, separated by an unstable steady state solution, as plotted in Fig. \ref{fig:sk3}.

\end{appendix}

\end{document}